\documentclass[nofootinbib,twocolumn,prd,preprintnumbers,superscriptaddress,aps]{revtex4-2}
\usepackage{amsmath}
\usepackage{amsfonts}
\usepackage{graphicx}
\usepackage[colorlinks=true,
linkcolor=blue,
citecolor=blue,
urlcolor=blue]{hyperref}
\usepackage{physics}
\usepackage{subcaption}
\usepackage{xcolor}
\usepackage{bbold}
\usepackage{cases}
\usepackage{braket}
\usepackage[T1]{fontenc}
\usepackage{mathrsfs}
\usepackage{enumerate}
\usepackage{bm}
\usepackage{multirow}
\usepackage{comment}
\usepackage{mathrsfs}

\numberwithin{equation}{section}

\begin{document}

\title{Coherent states in minimal-length Quantum Mechanics: inequivalent characterizations and emergent squeezing}

\author{Giuseppe Gaetano Luciano}
\email{giuseppegaetano.luciano@udl.cat}
\affiliation{Department of Chemistry, Physics and Environmental and Soil Sciences, Polytechnic School, University of Lleida, Av. Jaume II, 69, 25001 Lleida, Spain
}

\author{Pasquale Bosso}
\email{pasquale.bosso@ino.cnr.it}
\affiliation{CNR-INO, Istituto Nazionale di Ottica, Via Campi Flegrei 34, 80078 Pozzuoli, Italy}

\author{Daniel Chemisana}
\email{daniel.chemisana@udl.cat}
\affiliation{Department of Chemistry, Physics and Environmental and Soil Sciences, Polytechnic School, University of Lleida, Av. Jaume II, 69, 25001 Lleida, Spain
}

\date{\today}

\begin{abstract} 
Several approaches to quantum gravity suggest the emergence of a fundamental minimal length at the Planck scale. In quantum mechanics, this feature is naturally encoded through deformations of the Heisenberg algebra, leading to the Generalized Uncertainty Principle (GUP). While the phenomenological implications of GUP have been extensively explored, a consistent characterization of coherent states in minimal-length quantum mechanics remains elusive. In this work, we present a systematic analysis of coherent states for the one-dimensional harmonic oscillator. We show that the canonical equivalence among their standard characterizations - as eigenstates of the annihilation operator, displaced vacuum states and minimum-uncertainty wave packets - is generically lost in the presence of a minimal length. We then investigate the dynamical and semiclassical consequences of this inequivalence by comparing the evolution of generalized coherent states with that of states saturating the GUP. In particular, we demonstrate that minimal-length effects induce nontrivial deformations of phase-space trajectories and give rise to an intrinsic squeezing mechanism with no counterpart in ordinary quantum mechanics. These results establish a unified framework for coherence in GUP-based quantum theories and identify distinctive semiclassical signatures of minimal-length physics, opening a new avenue for probing quantum-gravitational effects.
\end{abstract}

\maketitle

\section{Introduction}
\label{Intro}
Recent insights from various candidates of quantum gravity, including String Theory, Loop Quantum Gravity, Quantum Geometry and Doubly Special Relativity, suggest that the interplay between quantum mechanics (QM) and general relativity (GR) necessitates a profound rethinking of the background geometry at the Planck scale. Specifically, these models converge on the idea that quantum spacetime is not a smooth continuous fabric as described by classical relativity but rather exhibits a complex structure characterized by a minimal measurable length~\cite{Hossenfelder:2012jw,Bosso:2023aht}. From a phenomenological perspective, an intriguing question involves understanding how the existence of such a fundamental scale might influence low-energy systems~\cite{Ashtekar:2002sn,Amelino-Camelia:2008aez,Addazi:2021xuf,AlvesBatista:2023wqm}.

A typical strategy to account for the finite spatial resolution in QM is to modify the commutator between the
position and momentum operators, leading to a Generalized Uncertainty Principle (GUP)~\cite{Kempf:1994su,Maggiore:1993kv,Das:2009hs}. For suitable deformations, the ensuing paradigm predicts a minimal position uncertainty, which fundamentally alters the structure of the phase space\footnote{In the following analysis, when we refer to a minimal length, we shall mean a minimal uncertainty in position. 
}. The most emblematic one-dimensional generalization of the Heisenberg commutator is the Kempf-Mangano-Mann model~\cite{Kempf:1994su}
\begin{equation}
    [x, p]=i\hbar f(p)\,,
    \label{GUP}
\end{equation}
where 
\begin{equation}
\label{KMMfp}
f(p)=\left(1+\beta p^2\right)\,,
\end{equation}
and $x$, $p$ denote position and momentum operators, respectively.
Furthermore, $\beta = \ell^2 \beta_0 / \hbar^2$, with $\ell$ being a reference length, commonly assumed to be the Planck length.
For positive values of the GUP parameter $\beta$, Eq.~\eqref{GUP} admits a minimal position uncertainty $\Delta x_{\text{min}} = \ell \sqrt{\beta_0}$. The standard QM is obtained for $\beta=0$, i.e., $f(p)=1$.

The GUP currently provides one of the most efficient frameworks for exploring the phenomenology of quantum gravity~\cite{Kempf:1994su,AlvesBatista:2023wqm,Das:2009hs,Scardigli:2014qka,Addazi:2021xuf,Buoninfante:2019fwr,Scardigli:2018jlm,Chemisana:2021xxz,Maggiore:1993kv,Scardigli:2003kr} (see Ref.~\cite{Bosso:2023aht} for a recent review). Despite extensive investigations in QM, particle physics and field theory, comparatively little attention has been directed toward its implications in the quasi-classical (decoherence) regime.
This domain is of particular theoretical and phenomenological significance, as it governs the emergence of classicality from quantum dynamics, and provides a critical testing ground for the consistency of semiclassical approximations under minimal-length deformations.
Furthermore, it may offer a viable setting for detecting imprints of quantum gravitational corrections at accessible energy scales, particularly in cosmological and astrophysical contexts~\cite{kiefer2012quantum}.

Important theoretical tools in quasi-classical quantum theory are the \emph{coherent states} (CSs).
Since they exhibit the most classical-like dynamics among quantum states, CSs provide a natural and intuitive framework for connecting QM with its classical limit. 
Various types of CSs have been studied so far. The Schr\"odinger-type minimum-uncertainty CS’s were originally defined by seeking solutions to the Schr\"odinger equation which~\cite{schrödinger1928collected} 
\begin{enumerate}[\hspace{0.5cm}a.]
    \item \it{minimize the Heisenberg uncertainty relation}.
    \label{property_a}
\end{enumerate}
These states represent the closest quantum analogs to classical trajectories, with balanced uncertainties in position and momentum.  In spite of their early identification, CSs did not attract full attention until Glauber established their pivotal role in the quantum theory of electromagnetic coherence~\cite{Glauber:1963tx}. In this context, CSs were formalized in the harmonic oscillator model as 
\begin{enumerate}[\hspace{0.5cm}a.]
    \setcounter{enumi}{1}
     \item \it eigenstates of the annihilation operator.
     \label{property_b}
\end{enumerate}
Such a property, which connects CSs to the quantized modes of the electromagnetic field, ensures that these states are applied across various fields, ranging from quantum optics~\cite{Glauber:1963tx}, where they can be used to describe the coherent laser light, to quantum information~\cite{Braunstein:2005zz}, where they serve as robust carriers of information. 

Building on Glauber's work, along with the Sudarshan's P-representation of quantum systems in phase space~\cite{Sudarshan:1963ts}, an alternative characterization of CSs was proposed as
\begin{enumerate}[\hspace{0.5cm}a.]
    \setcounter{enumi}{2}
     \item \it states that result from the application of a unitary ``displacement'' operator to the vacuum.  
     \label{property_c}
\end{enumerate}
This, in turn, amounts to the requirement that CSs be an infinite superposition of Fock (number) states with a Poisson distribution.
It is a matter of standard textbook calculations to demonstrate that the three properties outlined above are equivalent in QM, since each of them completely specifies the CS~\cite{cohen1977quantum}.

A preliminary construction of the CSs for the GUP-corrected harmonic oscillator was carried out within the KMM model~\cite{Nozari:2005it,Ghosh:2011ze,Bosso:2017ndq}. Using a perturbative approach, it was shown that the generalized CSs still satisfy all the known properties of the standard theory~\cite{Bosso:2017ndq}. On the other hand, by applying Klauder’s method~\cite{klauder1963continuous}, exact CSs were derived in Ref.~\cite{Pedram:2012ui}. However, the results indicate that the ordinary quantum description of laser light and the Poisson probability distribution of the CSs should be modified due to gravitational corrections. 

Recently, the explicit form of the CSs for the KMM model has been obtained in the momentum representation~\cite{Jizba:2022icu,Jizba:2023ygi}. This derivation shows that the GUP CSs coincide with the Tsallis probability amplitude, whose non-extensivity index $q$ monotonically increases with the GUP deformation parameter $\beta$ (see also~\cite{Luciano:2021ndh,Shababi:2020evc}). This insight offers a novel perspective for examining the transition from the quantum domain described by GUP to classical reality. Moreover, it paves the way for potential experimental investigations of GUP-based quantum gravitational phenomena through analog gravity models grounded in non-extensive thermostatistics.

Starting from the above premises, in this work we present a systematic study of GUP CSs. 
To this end, we build upon the results of Ref.~\cite{Bosso:2023nst}, which demonstrate that the only minimal-length deformed dynamics in one spatial dimension compatible with the relativity principle, parity invariance and a trivial composition of kinetic terms must be related to the standard formalism via a diffeomomorphism in momentum space, i.e. a canonical transformation.
We show that the equivalence between properties (\ref{property_a})-(\ref{property_c}) fails to hold when the canonical commutator is deformed according to Eq.~\eqref{GUP}. Furthermore, we derive the explicit form of the states that saturate the generalized uncertainty relation through a variational approach and compare them with the CSs defined as eigenstates of the generalized annihilation operator. We finally investigate the physical implications of this inequivalence by analyzing the dynamics, uncertainty evolution and squeezing properties of the resulting states, highlighting characteristic effects induced by the presence of a minimal length.

This work is organized as follows. In Sec.~\ref{QM}, we review the definition and main properties of CSs in standard QM. In Sec.~\ref{MLQM}, we extend the analysis to the case of GUP-based quantum theory, investigating how the deformation of the canonical commutator affects the equivalence between properties (\ref{property_a})-(\ref{property_c}). In Sec.~\ref{PIaD}, we discuss the physical implications of the formalism, with particular emphasis on the dynamics and uncertainty properties of the resulting states. Finally, conclusions and future perspectives are presented in Sec.~\ref{Conc}. The paper is complemented by three appendices containing additional computational details.

\section{Coherent states in QM}
\label{QM}

In this section, we review the conditions that characterize quasi-classical, or ``coherent,'' states in standard QM. To this end, we consider a quantum harmonic oscillator with mass \( m \), angular frequency \( \omega \) and
Hamiltonian $H=\dfrac{{p}^2}{2m} + \dfrac{1}{2}m\omega^2 {x}^2$, where \( x \) and \( p \) denote the position and momentum operators, respectively.

For this system, it is convenient to introduce the ladder (or annihilation and creation) operators, defined as
\begin{align}
    a &= \sqrt{\frac{m \omega}{2 \hbar}} \left( x + \frac{i}{m \omega} p \right),
    \label{a}\\[2mm]
    a^\dagger &= \sqrt{\frac{m \omega}{2 \hbar}} \left( x - \frac{i}{m \omega} p \right).
    \label{adag}
\end{align}
They obey the following algebra:
\begin{equation}
    [{x},{p}] = i \hbar
    \quad \Rightarrow \quad
    \begin{array}{c}
        [{a},{a}^\dagger] = 1,\\[3mm]
        [{a},{a}] = [{a}^\dagger,{a}^\dagger]=0.
    \end{array}
    \label{acom}
\end{equation}
Inverting Eqs.~\eqref{a}-\eqref{adag}, we are led to
\begin{equation}
\label{hatx}
     x =
    \sqrt{\frac{\hbar}{2 m \omega}} \left(a + a^\dagger\right),\quad \,\,
     p =
    i \sqrt{\frac{\hbar m \omega}{2}} \left(a^\dagger - a\right),
\end{equation}
with the Hamiltonian taking the well-known form
\begin{equation}
 H =\hbar \omega \left(a^\dagger a +\frac{1}{2}\right),
    \label{hamop}
\end{equation}
where the number operator $a^\dagger a$ counts the excitation quanta of the harmonic oscillator.

\subsection{Physical properties} 
\label{PP}

Building on the definitions introduced above, we now review the derivation of several characteristic properties of CSs, illustrating their mutual equivalence within the framework of ordinary QM.

\subsubsection{CSs as quasi-classical states}
 
Although CSs are genuine quantum states, one of their most remarkable features is that their dynamical evolution closely mimics that of a classical harmonic oscillator. 
This correspondence can be formally examined by considering the time evolution of the expectation values of the ladder operators through the Ehrenfest theorem, which can be written as
\begin{equation}
    i \hbar \frac{d}{dt}\langle a\rangle_{\psi}(t)=\langle [a,H]\rangle_\psi(t)\,,
    \label{expa}
\end{equation}
where the expectation values are computed with respect to a normalized (non-trivial) state $|\psi(t)\rangle$.

Since $[a,H]=\hbar\omega\, a$, we obtain
\begin{equation}
    i \frac{d}{dt}\langle  a\rangle_{\psi}(t)=\omega\,\langle  a\rangle_\psi(t)\,,
\end{equation}
which is solved by $\langle  a\rangle_{\psi}(t)=\langle  a\rangle_{\psi}(0)\,e^{-i\omega t}$. 
That is, the expectation value of the operator ${a}$ evolves as a phase.
Similarly, one can check that $\langle  a^\dagger\rangle_{\psi}(t)=\langle  a\rangle^*_{\psi}(0)\,e^{i\omega t}$.

Substituting these two relations into Eq.~\eqref{hatx}, we find
\begin{align}
   \langle  x\rangle_{\psi}(t) =&
   \sqrt{\frac{\hbar}{2 m \omega}} \left[\langle  a\rangle_\psi(0)\,e^{-i\omega t}+\langle a\rangle^*_\psi(0)\,e^{i\omega t}\right],
\label{Xexpval}\\[2mm]
   \langle  p\rangle_{\psi}(t) =&
   - i \sqrt{\frac{\hbar m \omega}{2}} \left[\langle  a\rangle_\psi(0)\,e^{-i\omega t}-\langle a\rangle^*_\psi(0)\,e^{i\omega t}\right].
    \label{Pexpval}
\end{align}
To ensure that $\langle x\rangle_{\psi}(t)$ and $  \langle p\rangle_{\psi}(t)$ match the classical
dynamics, the state $|\psi(t)\rangle$ must then satisfy \cite{cohen1977quantum}
\begin{equation}
\label{FirstCon}
    \langle  a\rangle_\psi(0)=\alpha_0\,,
\end{equation}
where $\alpha_0$ is a suitable constant characterizing the (dimensionless) complex amplitude of the classical oscillator.

A further condition on the state \(|\psi(t)\rangle\) follows from the observation that, 
in the classical limit - where 
\(\langle x^2 \rangle \simeq \langle x \rangle^2\) 
and 
\(\langle p^2 \rangle \simeq \langle p \rangle^2\) - %
the expectation value of the Hamiltonian, 
using Eqs.~\eqref{Xexpval} and~\eqref{Pexpval}, 
simplifies to
\begin{equation}
    \langle H \rangle_\psi = \hbar \omega |\alpha_0|^2\,,
\label{eqn:Hexpval}
\end{equation}
where the zero–point contribution $\hbar \omega/2$ has been neglected.
This approximation is valid in the semiclassical regime $\langle H \rangle_\psi \gg {\hbar\omega}/{2}$.

On the other hand, when the same expectation value is computed using Eq. \eqref{hamop}, we find
\begin{equation}
\label{exvalHam}
    \langle {H} \rangle_\psi
    = \hbar \omega \left( \langle {a}^\dagger {a} \rangle_\psi + \frac{1}{2} \right).
\end{equation}
Comparison with Eq.~\eqref{eqn:Hexpval} then gives, in the same semiclassical limit,
\begin{equation}
   \langle {a}^\dagger {a} \rangle_\psi \simeq |\alpha_0|^2,
    \label{SecCon}
\end{equation}

We shall see below that Eqs.~\eqref{FirstCon} and~\eqref{SecCon} uniquely determine the normalized vector $|\psi(0)\rangle$, up to an irrelevant phase factor. 
Thus, the parameter $\alpha_0$ that characterizes the classical motion of the oscillator at initial time plays a non-trivial role in defining the quantum CS.

\subsubsection{CSs as eigenstates of the operator $a$}
Let us define the operator $b(\alpha_0)$ as
\begin{equation}
     b(\alpha_0)= a-\alpha_0\,.
\end{equation}
To streamline the notation, henceforth we omit the $\alpha_0$-dependence of $b$. It is a straightforward check that 
\begin{equation}
      \langle b^\dagger b\rangle_\psi (0)=\langle a^\dagger a\rangle_\psi (0)-\alpha_0\langle  a^\dagger\rangle_\psi (0)-\alpha_0^*\langle  a\rangle_\psi (0)+|\alpha_0|^2\,.
 \end{equation}
 Using Eqs.~\eqref{FirstCon} and~\eqref{SecCon}, this can be simplified to $\langle b^\dagger b\rangle_\psi (0)=0$.  We then have 
 \begin{equation}
     \label{eigenstate}
b|\psi(0)\rangle=0\,\,\Longrightarrow\,\, a |\psi(0)\rangle=\alpha_0|\psi(0)\rangle,
 \end{equation}
which is property (\ref{property_b}) in Sec.~\ref{Intro}.
As is well known, by setting $\alpha_0 = 0$, one obtains a special CS, which is the ground state of the harmonic oscillator.

The other way around, if a given normalized vector obeys Eq.~\eqref{eigenstate}, the conditions~\eqref{FirstCon} and~\eqref{SecCon} are certainly satisfied. 
Therefore, we conclude that a normalized state associated with a classical dynamics of characteristic parameter $\alpha_0$ is a CS if and only if it is an eigenstate of $a$ with eigenvalue $\alpha_0$. 

Using the conventional notation, from now on we indicate $|\psi(0)\rangle\rightarrow|\alpha\rangle$, $\alpha_0\rightarrow\alpha$, so that Eq.~\eqref{eigenstate} takes the more compact form
\begin{equation}
\label{aCS}
 a |\alpha\rangle=\alpha|\alpha\rangle\,.
\end{equation}
We show below that this equation admits a unique solution up to a constant factor.

\subsubsection{CSs as Poisson superpositions of number states}

Let us first define the set of normalized eigenstates of the harmonic oscillator Hamiltonian, 
\(\{|n\rangle\}\), satisfying
\begin{equation}
    H |n\rangle = E_n |n\rangle, 
    \quad n \ge 0 \,,
\end{equation}
where \(E_n=\hbar \omega \left( n + \frac{1}{2} \right) \) denotes the energy spectrum of the oscillator. 
The states $\{|n\rangle\}$ form an orthonormal and complete basis of the Hilbert space, 
commonly referred to as the Fock space.

We now seek to determine the CS \(|\alpha\rangle\) introduced in Eq.~\eqref{aCS}. 
To this end, we expand it in the Fock space basis as
\begin{equation}
    |\alpha\rangle = \sum_{n=0}^{\infty} c_n(\alpha) \, |n\rangle \,,
    \label{CSexpansion}
\end{equation}
where \(c_n(\alpha)\) are complex coefficients to be determined.
From the definition of the ladder operator $a$, we get
\begin{equation}
     a|\alpha\rangle=\sum_n c_n(\alpha)\sqrt{n}|n-1\rangle\,.
\end{equation}
This can be inserted into Eq.~\eqref{aCS} to give the recurrence formula
\begin{equation}
    c_{n+1}(\alpha)=\frac{\alpha}{\sqrt{n+1}}\,c_n(\alpha)\,,
\end{equation}
or, equivalently, 
\begin{equation}
    c_n(\alpha)=\frac{\alpha^n}{\sqrt{n!}}\,c_0(\alpha)\,.
\end{equation}

For the sake of simplicity,
we set $c_0(\alpha)$ real and positive.
Imposing the normalization condition for $|\alpha\rangle$ and exploiting the orthonormality of $\{|n\rangle\}$, i.e., $\langle n|m\rangle=\delta_{n,m}$, we obtain
\begin{equation}
        \langle \alpha |\alpha\rangle=1 \,\,\,\Longrightarrow\,\,\,c_0^2(\alpha)\,e^{|\alpha|^2}=1\,,
\end{equation}
which implies
\begin{equation}
    c_n(\alpha)=\frac{\alpha^n}{\sqrt{n!}}\,e^{-\frac{|\alpha|^2}{2}}\,.
\end{equation}
The final expression of $|\alpha\rangle$ then reads
\begin{equation}
\label{infsup}
    |\alpha\rangle=e^{-\frac{|\alpha|^2}{2}}\sum_n\frac{\alpha^n}{\sqrt{n!}}|n\rangle\,.
\end{equation}
Assuming the oscillator is in the state $|\alpha\rangle$, the probability that the outcome of a measurement gives   $|n\rangle$ will be equal to
\begin{equation}
    \mathcal{P}_n(\alpha)=|\langle \alpha|n\rangle|^2=e^{-|\alpha|^2}\frac{|\alpha|^{2n}}{n!}\,,
\end{equation}
which is a \emph{Poisson distribution}. We conclude that, to obtain a quasi-classical state, we need to linearly superpose an infinite number of number states $|n\rangle$, each with probability $\mathcal {P}_n$.

\subsubsection{CSs as minimum uncertainty states}
Let us compute the position and momentum uncertainty of the quantum oscillator in the CS.
Following the definitions
in Eq. \eqref{hatx}, we are led to
\begin{align}
\label{valx}
    \langle  x\rangle_\alpha &= \sqrt{\frac{2\hbar}{m\omega}} \Re\{\alpha\}\,,\\[2mm]
    \Delta x_\alpha
    &\equiv \sqrt{\langle x^2\rangle_\alpha-\langle x\rangle_\alpha^2}= \sqrt{\frac{\hbar}{2m\omega}}, 
    \label{Dxalpha}\\[2mm]
\label{valp}
    \langle  p\rangle_\alpha &= \sqrt{2\hbar\hspace{0.2mm} m\hspace{0.2mm}\omega}\Im\{\alpha\}\,,\\[2mm]
    \Delta p_\alpha &\equiv \sqrt{\langle p^2\rangle_\alpha-\langle p\rangle_\alpha^2} = \sqrt{\frac{\hbar m \omega}{2}}.
    \label{Dpalpha}
\end{align}
By inserting the above expressions of $\Delta x_\alpha$ and $\Delta p_\alpha$ into the Heisenberg inequality, we obtain
\begin{equation}
\label{minHeis}
    \Delta x_\alpha\,\Delta p_\alpha = \frac{\hbar}{2}\,.
\end{equation}
Therefore, the state $|\alpha\rangle$ minimizes the Heisenberg uncertainty product [property (\ref{property_a}) in Sec.~\ref{Intro}].
This confirms that CS are the most ``classical'' among all quantum states.

\subsubsection{CSs from the vacuum}

We now consider the unitary operator 
\begin{equation}
\label{Dop}
     D(\alpha)=e^{\alpha a^\dagger-\alpha^* a}.
\end{equation}
Since the operators $\alpha a^\dagger$ and $\alpha^* a$ both commute with their commutators (see Eq.~\eqref{acom}), we can use 
Baker-Campbell-Hausdorff (BCH) formula
to write 
\begin{equation}
     D(\alpha)=e^{-\frac{|\alpha|^2}{2}}e^{\alpha a^\dagger}e^{-\alpha^* a}\,.
\end{equation}
Let $D(\alpha)$ act on the zero-particle state $|0\rangle$. Since
\begin{equation}
    e^{-\alpha^* a}|0\rangle=\sum_n\left(-1\right)^n\frac{\left(\alpha^* a\right)^n}{n!}|0\rangle=|0\rangle\,,
\end{equation}
we have
\begin{equation}
    D(\alpha)|0\rangle
    =e^{-\frac{|\alpha|^2}{2}}\sum_n\frac{\alpha^n}{\sqrt{n!}}|n\rangle\,.
    \label{Disp}
\end{equation}
However, from Eq.~\eqref{infsup}, we see that the r.h.s. is nothing but $|\alpha\rangle$. We thus infer that the CS is obtained through application of the ``displacement'' operator $D(\alpha)$ on the vacuum $|0\rangle$ [property (\ref{property_c}) in Sec.~\ref{Intro}].

\subsubsection{Position representation of CSs}
\label{Xrep}

The relation~\eqref{Disp} allows us to determine the wave function $\psi_\alpha(x)$ that characterizes the CS $|\alpha\rangle$ in the position representation $\{|x\rangle\}$.
Such a wavefunction is defined as
\begin{equation}
\label{mel}
    \psi_\alpha(x)\equiv \langle x|\alpha\rangle=\langle x |  D(\alpha)|0\rangle\,.
\end{equation}
To compute this matrix element, we rewrite the exponent in Eq.~\eqref{Dop} in terms of $x$ and $p$ as
\begin{equation}
    \alpha a^\dagger -\alpha^* a=\sqrt{\frac{m\omega}{\hbar}}\left(\frac{\alpha-\alpha^*}{\sqrt{2}}\right) x-\frac{i}{\sqrt{\hbar\hspace{0.2mm} m\hspace{0.2mm}\omega}}\left(\frac{\alpha+\alpha^*}{\sqrt{2}}\right) p\,,
\end{equation}
where we have used Eqs.~\eqref{a}-\eqref{adag}.
Applying again BCH formula, we acquire
\begin{equation}
     D(\alpha)=e^{\sqrt{\frac{m\omega}{\hbar}}\left(\frac{\alpha-\alpha^*}{\sqrt{2}}\right) x}\,e^{-\frac{i}{\sqrt{\hbar\hspace{0mm} m\hspace{0mm}\omega}}\left(\frac{\alpha+\alpha^*}{\sqrt{2}}\right) p}\,e^{\frac{{\alpha^*}^2-\alpha^2}{4}}\,.
\end{equation}
This can be inserted into Eq.~\eqref{mel} to give
\begin{equation}
    \psi_\alpha(x)=e^{\frac{{\alpha^*}^2-\alpha^2}{4}}\,e^{\sqrt{\frac{m\omega}{\hbar}}\left(\frac{\alpha-\alpha^*}{\sqrt{2}}\right)x}\langle x | e^{-\frac{i}{\sqrt{\hbar\hspace{0mm} m\hspace{0mm}\omega}}\left(\frac{\alpha+\alpha^*}{\sqrt{2}}\right) p} | 0\rangle\,.
    \label{psialpha}
\end{equation}

Let us now recall that $e^{-i\frac{\epsilon}{\hbar}p}$ applied on $|x\rangle$ translates this state by $\epsilon$ along the $x$-axis, i.e. $e^{-i\frac{\epsilon}{\hbar} p}|x\rangle=|x+\epsilon\rangle$. From Eq.~\eqref{psialpha}, we  get
\begin{align}
\nonumber    \psi_\alpha(x)&=e^{\frac{{\alpha^*}^2-\alpha^2}{4}}\,e^{\sqrt{\frac{m\omega}{\hbar}}\left(\frac{\alpha-\alpha^*}{\sqrt{2}}\right)x}\langle x-\sqrt{\frac{\hbar}{2m\omega}}\left(\alpha+\alpha^*\right)|0\rangle\\[2mm]
&=e^{\frac{{\alpha^*}^2-\alpha^2}{4}}\,e^{\sqrt{\frac{m\omega}{\hbar}}\left(\frac{\alpha-\alpha^*}{\sqrt{2}}\right)x}\psi_0\left(x-\sqrt{\frac{\hbar}{2m\omega}}\left(\alpha+\alpha^*\right)\right).
\end{align}
By using Eqs.~\eqref{valx} and~\eqref{valp}, this can be further manipulated to give
\begin{equation}
\psi_\alpha(x)=e^{i\theta_\alpha}\,e^{i\frac{\langle  p\rangle_\alpha}{\hbar}x}
    \psi_0\left(x-\langle  x\rangle_\alpha\right)\,,
\end{equation}
where we have defined the real factor
\begin{equation}
    \theta_{\alpha}\equiv - \frac{\langle  x\rangle_\alpha\langle p\rangle_\alpha}{2\hbar}\,.
\end{equation}
Therefore, the wave function $\psi_\alpha(x)$ of the CS of the oscillator is obtained by shifting the wave function of its ground state by the quantity $\langle x\rangle_\alpha$ along the $x$-axis and multiplying it by the oscillating exponential $e^{i\frac{\langle  p\rangle_\alpha}{\hbar}x}$. 

By using the explicit form of the wave function of the ground state, 
\begin{equation}
    \psi_0(x)=\left(\frac{m\omega}{\pi\hbar}\right)^{\frac{1}{4}}e^{-\frac{m\omega x^2}{2\hbar}}\,,
\end{equation}
we finally obtain
\begin{equation}
    \psi_\alpha(x)=
    \left(\frac{m\omega}{\pi\hbar}\right)^{\frac{1}{4}}
e^{i\theta_\alpha}\hspace{0.2mm}e^{i\frac{\langle p\rangle_\alpha}{\hbar}x}\,
    e^{-\left(\frac{x-\langle x\rangle_\alpha}{2\Delta x_\alpha}\right)^2}\,,
    \label{psialphafinal}
\end{equation}
where we have used Eq.~\eqref{valx}. 

In turn, the form of the wave packet associated with $|\alpha\rangle$ is 
\begin{equation}
\label{Gausx}
    |\psi_\alpha(x)|^2=\sqrt{\frac{m\omega}{\pi\hbar}}e^{-\frac{1}{2}\left(\frac{x-\langle x\rangle_\alpha}{\Delta x_\alpha}\right)^2}\,,
\end{equation}
which is a Gaussian profile, consistent with the property that the product $\Delta x_\alpha\,\Delta p_{\alpha}$ in the CS is always minimal.

It should be noted that the same Gaussian structure can be obtained by applying Schr\"odinger's method for minimizing the uncertainty relation, which is based on solving the differential equation
~\cite{cohen1977quantum,Jizba:2022icu}
\begin{equation}
\label{difeq}
    \left(O_2-i\gamma\hspace{0.2mm} O_1\right)|\widetilde \psi\rangle=0\,,
\end{equation}
where $O_1=x-\langle x\rangle$ and $O_2=p-\langle p\rangle$, along with the condition $\gamma=\frac{i\langle[O_2,O_1]\rangle_{\widetilde\psi}}{2\langle  O^2_1\rangle_{\widetilde\psi}}=\frac{\hbar}{2\Delta x^2_{\widetilde\psi}}$.
Expanding Eq.~\eqref{difeq} in the position representation gives 
\begin{equation}
    \frac{d\widetilde\psi(x)}{dx}+\frac{\gamma}{\hbar}\left[(x-\langle x\rangle_{\widetilde\psi})-i\frac{\langle p\rangle_{\widetilde\psi}}{\gamma}\right]\widetilde\psi(x)=0\,,
\end{equation}
whose solution is
\begin{equation}
\label{psit}
    \widetilde\psi(x)= C\hspace{0.2mm}e^{i\frac{\langle  p\rangle_{\widetilde\psi}}{\hbar}x}\hspace{0.2mm}e^{-\frac{\gamma}{2\hbar}\left(x^2-2x\langle x\rangle_{\widetilde\psi}\right)}\,.
\end{equation}
Imposing normalization, we obtain
\begin{equation}
    C=\left(\frac{\gamma}{\pi\hbar}\right)^{\frac{1}{4}}e^{-\frac{\gamma}{2\hbar}\langle x \rangle_{\widetilde\psi}^2}\,,
\end{equation}
which can be substituted into Eq.~\eqref{psit} to yield
\begin{equation}
     \widetilde\psi(x)= \bigg(\frac{1}{2\pi\Delta x^2_{\widetilde\psi}}\bigg)^{\frac{1}{4}}e^{i\frac{\langle p\rangle_{\widetilde\psi}}{\hbar}x}\hspace{0.2mm}e^{-\frac{1}{4\Delta x^2_{\widetilde\psi}}\left(x- \langle x \rangle_{\widetilde\psi}\right)^2}\,.
\end{equation}
The latter coincides (up to an irrelevant global phase) with Eq.~\eqref{psialphafinal} if $\Delta x_{\widetilde\psi}$ is set equal to the minimum position uncertainty of the ground state of the harmonic oscillator, i.e. $\Delta x_{\widetilde\psi}=\sqrt{h/(2m\omega)}$.

\subsubsection{Momentum representation of CSs}

For later convenience, it is also useful to determine the CS in the momentum representation. This is obtained by Fourier-transforming the position-space wave function, i.e.,
\begin{equation}
\label{momrep}
    \phi_\alpha(p)\equiv \langle p|\alpha\rangle = \frac{1}{\sqrt{2\pi\hbar}}\int_{-\infty}^{\infty}e^{-i\frac{p x}{\hbar}}\psi_\alpha(x)\hspace{0.2mm}dx\,,
    \end{equation}
where we have used the spectral resolution of the identity in the position space, along with the Fourier factor $\langle p|x\rangle=e^{-i\frac{px}{\hbar}}/\sqrt{2\pi\hbar}$.

Substituting Eq.~\eqref{psialphafinal} into Eq.~\eqref{momrep} and integrating, we obtain
\begin{equation}
    \phi_\alpha(p)=\left(\frac{1}{\pi\hbar m\omega}\right)^{\frac{1}{4}}e^{-i\theta_\alpha}\hspace{0.2mm}e^{-i\frac{\langle  x\rangle_\alpha}{\hbar}p}\hspace{0.2mm}e^{-\left(\frac{p-\langle  p\rangle_\alpha}{2\Delta p_\alpha}\right)^2}\,,
    \label{csmomrep}
\end{equation}
where we have employed Eq.~\eqref{minHeis}. We remark that the same result is obtained by solving Eq.~\eqref{difeq} in the momentum representation.

\subsection{Mathematical properties}

Since CSs are eigenstates of the non-Hermitian operator $a$, it is not true a priori they satisfy orthogonality and closure relations. To verify the former, let us evaluate the scalar product $\langle \alpha|\alpha'\rangle$. Using Eq.~\eqref{infsup}, we acquire 
\begin{equation}
\label{ort}
    \langle \alpha|\alpha'\rangle 
    \neq \delta(\alpha - \alpha')\,.
\end{equation}
This implies that CSs are not orthogonal, which is linked to the fact that they are eigenvectors of the non-self-adjoint annihilation operator $a$.

Furthermore, one can check that CSs satisfy the closure relation
\begin{equation}
\label{res}
\frac{1}{\pi}\int |\alpha\rangle\langle \alpha|\,d\alpha=\mathbb{1}\,,
\end{equation}
where the integral is performed over the entire complex $\alpha$ plane, i.e. $d^2\alpha\equiv d\Re{\alpha}\,d\Im{\alpha}$. Therefore 
such states form an overcomplete basis. We remark that this property underlies the Glauber–Sudarshan $P$-representation~\cite{Sudarshan:1963ts,Glauber:1963tx}, which plays a central role in quantum optics by providing a convenient framework for representing quantum states of the electromagnetic field~\cite{schleich2011quantum}.

\section{Minimal-length QM}
\label{MLQM}

One of the primary research areas in quantum gravity phenomenology focuses on minimal-length models. In fact, combining principles from GR and quantum theory heuristically suggests the emergence of a fundamental minimal-length at Planck scale~\cite{Amati:1987wq,Gross:1987kza,Amati:1988tn,Konishi:1989wk,Maggiore:1993kv,Capozziello:1999wx,Kempf:1994su,Scardigli:1999jh,Adler:2001vs,Hossenfelder:2012jw,Iorio:2022ave}. For instance, during scattering processes with extremely high center-of-mass energy, black holes can form at sufficiently small impact parameters, preventing the resolution of distances below a certain scale~\cite{Mead:1964zz,Scardigli:1999jh}. This idea is independently substantiated by various quantum gravity theories, including string theory~\cite{Gross:1987ar}, loop quantum gravity~\cite{Rovelli:1994ge}, asymptotic safety~\cite{Ferrero:2022hor}, etc. (see~\cite{Hossenfelder:2012jw} for a recent review).

In the framework of non-relativistic QM,  introducing a minimal-length scale is commonly achieved by modifying the Heisenberg algebra, which results in a generalized uncertainty principle (GUP)~\cite{Kempf:1994su}. The ensuing models incorporate the minimal-length scale through the Robertson-Schr\"odinger (RS) relation~\cite{Robertson,Schrodinger:1930}, which sets a fundamental limit on localization.
Although in the following we will exclusively consider this quantum-mechanical treatment, it is worth mentioning that the same description percolates into quantum field theory implying specific commutation relations between fields \cite{Bosso:2024nmn}.

A one-dimensional generalization of the Heisenberg commutator \eqref{acom} can be expressed as in Eq.~\eqref{GUP}, here rewritten for convenience, 
\begin{equation*}
    [x,  p] = i \hbar f(p)\,,
\end{equation*}
where $f(p)$ is a suitable real-valued function. Standard QM is naturally recovered when $f(p)$ is the identity. 
From the above commutator, the RS inequality gives the generalized uncertainty relation
\begin{equation}
    \Delta x\, \Delta p
    \ge
    \frac{\hbar}{2} |\langle f(p)\rangle|.
    \label{RS}
\end{equation}

As a paradigmatic example, the KMM model is built upon the function $f(p)$ in Eq.~\eqref{KMMfp}~\cite{Kempf:1994su}. In this setting, the inequality Eq.~\eqref{RS} yields
\begin{align}
        \nonumber
        \Delta x\, \Delta p&\ge\frac{\hbar}{2}\left(1+\beta\langle p^2\rangle\right)\\[2mm]
        &=\frac{\hbar}{2}\left\{1+\beta\left[(\Delta p)^2+\langle p\rangle^2\right]\right\}.
        \label{KMMine}
\end{align}
One can read off the minimal position uncertainty
\begin{equation}
    \Delta x_{\text{min}}=\hbar\sqrt{\beta\left(1+\beta\langle p\rangle^2\right)}\,,
    \label{Dxmin}
\end{equation}
which reduces to $\Delta x_{\text{min}} = \hbar \sqrt{\beta} = \ell \sqrt{\beta_0}$ for states $\psi$ satisfying $\langle p\rangle_{\psi}=0$. Here, the minimal length scale $\ell$ is defined below Eq.~\eqref{KMMfp}.
In passing, we note that alternative GUP models, involving either negative values of $\beta$~\cite{Magueijo:2001cr,Jizba:2009qf,Buoninfante:2019fwr,Ong:2018zqn} or more exotic functions $f(x, p)$~\cite{Tawfik:2014zca,Vagenas:2019wzd,Mignemi:2009ji,Dago, Snyder:1946qz,Ivetic:2026rxo}, have also been considered in the literature. 

To maintain the analysis at a general level, in what follows we will consider a GUP model characterized by a positive function $f(p)$, the specific form of which will be provided only in the final stage. To simplify calculations, let us introduce the auxiliary operator $q$ such that~\cite{Bosso:2021koi,Bosso:2023nst,Bosso:2024nmn}
\begin{equation}
\label{canpq}
    [q, p]=i\hbar\,.
\end{equation}
Assuming that $x\equiv x(q, p)$ is differentiable, the relation between $x$ and $q$ can be found by noting that 
\begin{equation}
    [x(q, p), p]=[q,p]\hspace{0.2mm}{\frac{\partial x(q,p)}{\partial q}}=i\hbar{\frac{\partial x(q,p)}{\partial q}}\,.
\end{equation}
From comparison with Eq.~\eqref{GUP}, we can easily see that
\begin{equation}
   {\frac{\partial x (q,p)}{\partial q}}=f(p)\,. 
\end{equation}
As shown in \cite{Bosso:2021koi}, 
it is convenient to employ a symmetric ordering prescription in defining the operator ${x}$. Thus, we can write
\begin{equation}
    x
    = \frac{1}{2} \left\{f(p),q\right\},
    \label{factord}
\end{equation}
where the symbol $\{\cdot,\cdot\}$ stands for anti-commutation.
The expression above, among other things, allows for the ordinary measure for momentum space.
However, other ordering prescriptions can be applied in an equivalent way and mapped to each other \cite{Bosso:2021koi}.
For concreteness, in the representation where $p$ is multiplicative, we have 
\begin{equation}
\label{sympre}
      x
      = \frac{i\hbar}{2} \left\{f(p),\frac{d}{dp}\right\}.
\end{equation}

Similarly, one can introduce the wavenumber operator $k$ such that
\begin{equation}
    [x, k]=i\,.
    \label{waven}
\end{equation}
Assuming that $p = p (k)$ is differentiable and following the same reasoning as above, we are led to
\begin{equation}
    [x, p(k)] 
    = i {\frac{\partial p(k)}{\partial k}}
    \,\,\,\Longrightarrow\,\,\, {\frac{\partial p(k)}{\partial k}} 
    = \hbar f(p)\,.
    \label{pkrel}
\end{equation}
It is worth noting that, for a minimal length to be present, the wavenumber operator must be bounded~\cite{Bosso:2023sxr}. Therefore, unlike $p$, the operator $k$ cannot obey a linear composition law.

An important remark is in order here. Since both pairs $(q,p)$ and $(x,k)$ are canonical, 
the transformation between them constitutes a canonical map. As discussed in 
Ref.~\cite{Bosso:2023nst}, working with the pair $(q,p)$ yields a Hamiltonian that 
is formally equivalent to that of ordinary Galilean quantum theory, with the 
important \emph{caveat} that $q$ does not represent the physical position 
operator. Following this prescription, the dynamics, for example, of the 
harmonic oscillator within the GUP framework must be analyzed by considering the Hamiltonian
\begin{equation}
    \mathcal{H} = \frac{p^{2}}{2m} + \frac{1}{2}\, m \omega^{2} q^{2}\,,
    \label{GUPHam}
\end{equation}
together with the commutation relation~\eqref{canpq}. It is implicitly understood that ${\mathcal{H}}$ can be re-expressed in terms of ${x}$ using Eq.~\eqref{factord}. Furthermore,  Eq.~\eqref{GUPHam} represents the only physically consistent Hamiltonian within the GUP framework.
Any alternative choice would be incompatible with invariance under GUP-deformed Galilean transformations~\cite{Bosso:2023nst}. 
This does not imply that the resulting theory is trivial.
Indeed, although the spectrum of Eq.~\eqref{GUPHam} is identical to that of ordinary QM, the dynamics expressed in terms of the physical position ${x}$ is clearly deformed.

As a consequence of the formal equivalence between the ordinary and deformed Hamiltonians, and
following Eqs.~\eqref{a} and~\eqref{adag}, we introduce the new ladder operators
\begin{align}
    {a}_f
    =& \sqrt{\frac{m \omega}{2 \hbar}} \left(q + \frac{i}{m \omega}  p\right),
\label{d}\\[2mm]
    {a}_f^\dagger
    =& \sqrt{\frac{m \omega}{2 \hbar}} \left(q - \frac{i}{m \omega} p\right),
    \label{ddag}
\end{align}
where the subscript ``$f$'', although purely formal, indicates that the corresponding quantity refers to the generalized quantum-mechanical framework defined by Eq.~\eqref{GUP}.

With the aid of Eq.~\eqref{canpq}, it is straightforward to verify that $[{a}_f, {a}_f^\dagger] = 1$, while all other commutators vanish. In terms of such operators, the Hamiltonian~\eqref{GUPHam} can be factorized in the standard form
\begin{equation}
     H=\hbar \omega \left({a}_f^\dagger {a}_f +\frac{1}{2}\right).
\end{equation}

In the sense outlined in Sec.~\ref{QM}, we define the CS 
$|\alpha\rangle_f$ associated with the algebra  \eqref{d}–\eqref{ddag} as the state satisfying
\begin{equation}
    {a}_f |\alpha\rangle_f
    = \alpha_f|\alpha\rangle_f,
\label{AnnGUP}
\end{equation}
i.e. the CS defined in this way is still eigenstate of the annihilation operator in Eq.~\eqref{d}.
It is worth emphasizing that, although $|\alpha\rangle_f$ is formally identical to the CS $|\alpha\rangle$ introduced in Sec.~\ref{QM}, this analogy holds as long as the ladder operators are defined in terms of the pair of canonical operators $(q,  p)$ as in Eqs.~\eqref{d} and~\eqref{ddag}, while it breaks down for other definitions~\cite{Pedram:2012ui,Nozari:2005it}.

Similarly, by defining the set of Fock states 
$\{|n\rangle_f\}$ ($n\ge0$) in such a way that
\begin{align}
    {a}_f|n\rangle_f
    &= \sqrt{n} |n-1\rangle_f,\\[2mm] 
    {a}_f^\dagger|n\rangle_f
    &= \sqrt{n+1} |n+1\rangle_f,
\end{align}
one can formally write the CS $|\alpha_f\rangle$ as a Poisson distribution of Fock states according to
\begin{equation}
\label{infsupGUP}
    |\alpha\rangle_f
    = e^{-\frac{|\alpha_f|^2}{2}} \sum_n \frac{\alpha_f^n}{\sqrt{n!}} |n\rangle_f.
\end{equation}

Following Eq.~\eqref{Dop}, let us now define the generalized displacement operator ${D}_f(\alpha_f) = e^{\alpha_f {a}_f^\dagger - \alpha_f^* {a}_f}$.
As in the ordinary case, the application of ${D}_f$ on the vacuum $|0\rangle_f$ gives
\begin{equation}
    {D}_f(\alpha_f) |0\rangle_f
    = |\alpha\rangle_f.
    \label{DispGUP}
\end{equation}
In other terms, the action of ${D}_f(\alpha_f)$ on the vacuum $|0\rangle_f$ displaces it into the generalized CS $|\alpha\rangle_f$. 

In turn, Eq.~\eqref{DispGUP} allows us to find the expression of $|\alpha\rangle_f$ in the representation of the auxiliary position $\{|q\rangle\}$.
Retracing the same computations as in Sec.~\ref{PP} leads us to
\begin{align}
\nonumber
    \psi_{\alpha_f}(q)
    &= \langle q |  D(\alpha_f)|0\rangle_f\\[2mm]
    &= \left(\frac{m\omega}{\pi\hbar}\right)^{\frac{1}{4}} e^{i\theta_{\alpha_f}} e^{i\frac{\langle  p\rangle_{\alpha_f}}{\hbar}q} e^{-\left(\frac{q-\langle q\rangle_{\alpha_f}}{2\Delta q_{\alpha_f}}\right)^2},
    \label{qrep}
\end{align}
where $\theta_{\alpha_f}\equiv-\dfrac{\langle  q\rangle_{\alpha_f}\langle p\rangle_{\alpha_f}}{2\hbar}$, $\langle q\rangle_{\alpha_f}$ and $\langle p\rangle_{\alpha_f}$ are the expectation values of the auxiliary position ${q}$ and the momentum ${p}$ on a generalized CS $|\alpha\rangle_f$, whose values in terms of the eigenvalue $\alpha_f$ are
\begin{align}
    \langle {q} \rangle_{\alpha_f}
    &= \sqrt{\frac{2 \hbar}{m \omega}} \, \Re\{\alpha_f\},\\[2mm]
    \langle {p} \rangle_{\alpha_f}
    &= \sqrt{2 \hbar m \omega} \, \Im\{\alpha_f\},
\end{align}
and where $\Delta q_{\alpha_f}$ is the uncertainty in ${q}$, given by $\Delta q_{\alpha_f} = \sqrt{\frac{\hbar}{2 m \omega}}$.

Therefore, Eqs.~\eqref{AnnGUP} and~\eqref{DispGUP} demonstrate that the CS
still satisfies the properties (\ref{property_b}) and (\ref{property_c}) outlined in Sec.~\ref{Intro}. This is equivalent to saying that these states 
can be expressed as a Poissonian superposition of the harmonic oscillator 
eigenstates $\{|n\rangle_f\}$ and can be obtained from the vacuum state $|0\rangle_f$ by the action of the 
displacement operator.
Moreover, Eq.~\eqref{qrep} indicates that it admits the same formal expression as the standard CS in the auxiliary position representation.

These features were to be expected,  as the aforementioned properties  depend only implicitly on the position and momentum through the definitions Eqs.~\eqref{d} and~\eqref{ddag} of the ladder operators. In fact, when dealing with the auxiliary position $q$, the ladder operators and the Hamiltonian are formally identical to their counterparts in QM (see Eqs.~\eqref{a},~\eqref{adag} and~\eqref{hamop}), thus ``hiding'' the effects of the minimal length. 

\subsection{Minimizing the GUP}
\label{Minimize}
It is now interesting to investigate how the property (\ref{property_a}) of minimization of the uncertainty relation is modified. Clearly, as long as the pair \(({q}, {p})\) is considered, Eq.~\eqref{minHeis} remains valid, that is, \(\Delta q_\alpha \Delta p_\alpha = \hbar/2\).
However, when the operator ${x}$ is taken into account, the situation becomes considerably more intricate.

In what follows, when no confusion arises, we will drop the subscript $f$ in order to simplify the notation.

\begin{figure}[t]
\centering\includegraphics[width=0.47\textwidth]{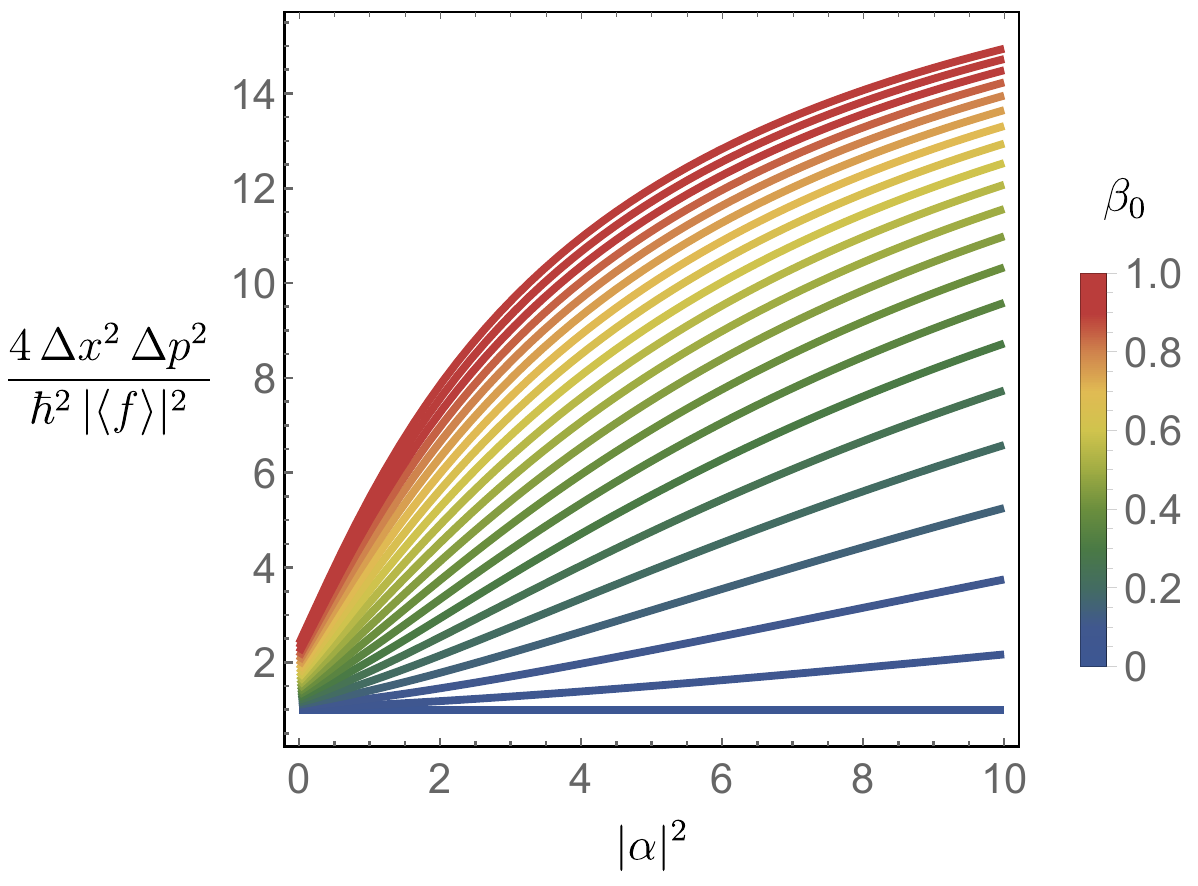}
\caption{
Plot of the ratio between the uncertainty product $\Delta x^2_{\alpha} \Delta p^2_{\alpha}$ and its theoretical minimum $\frac{\hbar^2}{4} |\langle f( p)\rangle_{\alpha}|^2$ as a function of $|\alpha|^2$ and for different values of $\beta_0$. We have set $\Re\{\alpha\} = \Im\{\alpha\}$ and used natural units.
}%
\label{Fig0}%
\end{figure}

As a specific example, let us focus on the case of the KMM model, considering generalized CS $|\alpha\rangle$ such that $\langle  q\rangle_{\alpha}=\langle  p\rangle_{\alpha}=0$, which in turn imply $\Re\{\alpha\}=\Im\{\alpha\}=0$.
Then, it is easy to show, while the $q-p$ uncertainty relation is indeed saturated,  for the $x-p$ uncertainty relation we get
\begin{equation}
    \Delta x^2_{\alpha} \Delta p^2_{\alpha}
    = \frac{\hbar^2}{4} + \frac{\beta\hbar^3 m \omega}{4} + \frac{7\beta^2 h^4 m^2\omega^2}{16}.
    \label{DxDp}
\end{equation}
On the other hand, when the minimal uncertainty product is computed for these states according to Eq. \eqref{RS}, we find
\begin{equation}
    \frac{\hbar^2}{4} |\langle f( p)\rangle_{\alpha}|^2
    = \frac{\hbar^2}{4} + \frac{\beta h^3 m\omega}{4} + \frac{\beta^2 h^4 m^2\omega^2}{16}.
    \label{RightSide}
\end{equation}
The difference between these two expressions shows that the definition~\eqref{AnnGUP} of generalized CSs is, in general, not consistent with the saturation of the RS inequality~\eqref{RS}. This remains true for any value of \( \alpha \), as illustrated in Fig.~\ref{Fig0}. However, the two expressions~\eqref{DxDp} and~\eqref{RightSide} coincide up to first order in \( \beta \).

At this point, it should be noted that if the GUP~\eqref{KMMfp} is regarded as a first-order perturbative approximation to a more general uncertainty relation - which, in general, involves higher-order powers of~$\beta$~\cite{GreenSchwarzWitten1987,Amati:1988tn} - then, for consistency, only terms up to linear order in~$\beta$ should be retained also in Eqs.~\eqref{DxDp} and~\eqref{RightSide}.
In this case, the CS defined in Eq.~\eqref{AnnGUP} likewise saturate the uncertainty relation, and the equivalence among properties (\ref{property_a})-(\ref{property_c}) remains valid even within the GUP framework.

On the other hand, if one adopts the standpoint that the GUP~\eqref{KMMfp} represents an exact model from the outset~\cite{Kempf:1994su}, the aforementioned equivalence no longer holds, thereby raising the question of which states saturate the inequality~\eqref{KMMine}.
To determine such states, it is important to recall that the standard Heisenberg minimization procedure is generally inapplicable when the commutator \([ {x}, {p} ]\) is a \( q \)-number, as occurs in the presence of the GUP.
In this case, an appropriate variational approach must instead be employed~\cite{Jackiw:1968zzb} (see also the recent analysis in the context of relativistic corrections to the Heisenberg principle~\cite{Luciano:2026zke}). 

Specifically, given an arbitrary function $f({p})$ in Eq.~\eqref{GUP} compatible with a minimal length, we shall initially
set the expectation values of $x$ and $p$ to zero. The more general case, in which these quantities assume non-vanishing mean values, will be addressed later.
Under these assumptions, the variational approach of Ref.~\cite{Jackiw:1968zzb} leads to the differential equation
\begin{equation}
\label{GenDifEq}
    \left[\frac{x^2}{A^2} + \frac{p^2}{B^2} - \frac{2 f({p})}{\langle f({p}) \rangle}_{\text{sat}}\right] |\phi_{\text{sat}}\rangle=0,
\end{equation}
where the quantities $A$ and $B$ are constrained to the position and momentum by the following relations
\begin{align}
    A &= \Delta x_{\text{sat}},&
    B &= \Delta p_{\text{sat}}\,,
\end{align}
with $\Delta x_{\text{sat}}$ and $\Delta p_{\text{sat}}$ saturating the uncertainty relation, that is
\begin{equation}
\label{dxsat}
    \Delta x_{\text{sat}} \Delta p_{\text{sat}}
    = \frac{\hbar}{2} \langle f({p}) \rangle_{\text{sat}}.
\end{equation}
Simple algebra allows us to rewrite Eq. \eqref{GenDifEq} in the equivalent form
\begin{equation}
    \left(\frac{x}{\Delta x_{\text{sat}}} - i \frac{p}{\Delta p_{\text{sat}}}\right) \left(\frac{x}{\Delta x_{\text{sat}}} + i \frac{p}{\Delta p_{\text{sat}}}\right) |\phi_{\text{sat}}\rangle = 0.
    \label{Jack}
\end{equation}
It is worth observing that, when $\Delta x_{\text{sat}}$ and $\Delta p_{\text{sat}}$ acquire the values in Eqs. \eqref{Dxalpha} and \eqref{Dpalpha}, the two terms in brackets in Eq. \eqref{Jack} correspond to the creation and annihilation operators of Eqs. \eqref{a} and \eqref{adag}.
Thus, in ordinary QM, that is when $f(p) = 1$, Eq. \eqref{Jack} is trivially solved by the vacuum state.
Alternatively, if the values of the position and momentum uncertainties are left undefined, with the constraint that the state saturates the Heisenberg uncertainty relation, the equation above describes squeezed vacuum states.
However, in minimal-length QM, the same two terms do not constitute suitable ladder operators, as it can be easily checked comparing Eq. \eqref{Jack} with Eqs. \eqref{d} and \eqref{ddag}.

By using Eqs. \eqref{waven} and \eqref{pkrel}, we can solve the differential equation \eqref{Jack} in momentum space, obtaining (see Appendix \ref{Momrepres})
\begin{equation}
    \phi_{\text{sat}}(p)
    = \frac{c_2}{\sqrt{f(p)}} \exp \left[- \frac{\hbar \langle f(p)\rangle_{\text{sat}}}{2 \Delta p^2_{\text{sat}}} \int_0^p \frac{p'}{\hbar f(p)}  \, d p'\right].
\end{equation}
Considering the specific case of the deformation introduced in Eq. \eqref{KMMfp}, we then find (up to a global phase)
\begin{equation}
    \phi_{\text{sat}}(p)
    = c \left(1 + \beta p^2\right)^{-\frac{3}{4} - \frac{1}{4\beta \Delta p^2_{\text{sat}}}},
    \label{GUPCSbetarid}
\end{equation}
with
\begin{equation}
    c
    = \left(\frac{\beta}{\pi}\right)^{\frac{1}{4}} \sqrt{\frac{\Gamma\left[\frac{1}{2}\left(3+\frac{1}{\beta \Delta p^2_{\text{sat}}}\right)\right]}{\Gamma\left(1+\frac{1}{2\beta \Delta p^2_{\text{sat}}}\right)}},
    \label{norm}
\end{equation} 
where $\Gamma(z)$ is the gamma function.
With the above choice of integration constant, one can show that Eq.~\eqref{GUPCSbetarid} properly reduces to Eq.~\eqref{csmomrep} (evaluated for $\langle {x} \rangle_{\alpha} = \langle {p} \rangle_{\alpha} = 0$) in the limit $\beta \to 0$.

\begin{figure}[t]
\centering\includegraphics[width=0.45\textwidth]{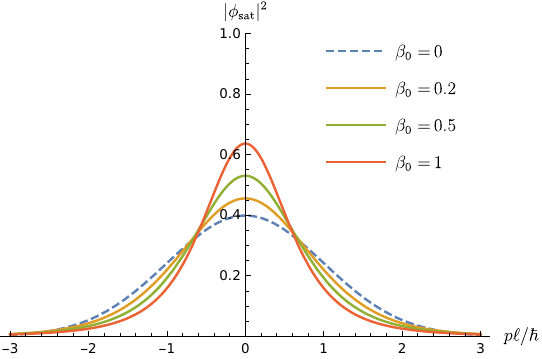}
\caption{Plot of $|\phi_{\text{sat}}|^2$ as a function of the dimensionless momentum $p \ell / \hbar$, for various values of the GUP parameter $\beta_0$.
We set $\Delta p_{\text{sat}}= \hbar / \ell$.}%
\label{Fig1}%
\end{figure}

The consistency in the vanishing $\beta$ limit is clearly illustrated in Fig.~\ref{Fig1}, 
where the function $|\phi_{\text{sat}}|^2$ is plotted versus $p$ for different values of $\beta$. 
As expected, the curves gradually converge toward the HUP result (black dashed line) as the GUP parameter $\beta$ decreases. 
Conversely, quantum-gravitational effects manifest as a momentum-space probability density that becomes increasingly narrow 
and sharply peaked around $p = 0$, with the strength of this effect growing for larger values of $\beta$. The reversed behavior is exhibited in the position-space representation, where increasing $\beta$ leads instead to a broader probability distribution (see the discussion in Appendices~\ref{AppA} and~\ref{AppB}).


It is worth pointing out that the maximally-localized states described in \cite{Kempf:1994su} belong to the family of states presented here.
In fact, imposing $\Delta p_{\text{sat}} = 1/\sqrt{\beta}$, so to achieve a minimal uncertainty in position, we obtain
\begin{equation}
    \phi_{\text{sat}} (p)
    = \sqrt{\frac{2 \sqrt{\beta}}{\pi}}
    \frac{1}{1 + \beta p^2}\,,
\end{equation}
which indeed corresponds to the maximally-localized states of \cite{Kempf:1994su} once the different ordering is accounted for according to \cite{Bosso:2021koi}.

Concluding this section, we note an issue arising from the states in Eq. \eqref{GUPCSbetarid}. Specifically, let us consider their behavior in the regime $p \gg 1/\sqrt{\beta}$. In this limit, we obtain
\begin{equation}
\phi_{\text{sat}} (p) \propto p^{-\frac{3}{2} - \frac{1}{2 \beta \Delta p_{\text{sat}}^2}}.
\end{equation}
It is therefore clear that, for a given value of $\beta$, the expectation values of arbitrary powers of $p$ are not always well defined. To see this, let us consider the expectation value of $p^n$. For large momenta, the corresponding integrand over $p \in \mathbb{R}$ behaves as
\begin{equation}
\phi_{\text{sat}}^* p^n \phi_{\text{sat}}
\sim
p^{n-3 - \frac{1}{\beta \Delta p_{\text{sat}}^2}}.
\end{equation}
Thus, expectation values of powers of the momentum with $n \geq 2 + \frac{1}{\beta \Delta p_{\text{sat}}^2}$ are not well defined. Consider, for example, the model presented in \cite{Kempf:1994su}. In this case, expectation values of powers of $p$ with $n \geq 3$ diverge. As we shall see, this poses a problem for the computation of energy uncertainties for such states.

\subsection{Displaced minimal uncertainty states}
So far, we have focused on states with vanishing expectation values of position and momentum. We now consider the case in which the position has a non-zero expectation value, which can be realized by translating the state in space by the corresponding amount.
To this end, we recall that, in ordinary QM, spatial translations are generated by the momentum operator. Specifically, the translation operator by a displacement \(a\) is given by
${T}(a) = e^{-i a {p} / \hbar} = e^{-i a {k}}$, where $\hbar k= p$~\cite{cohen1977quantum}.

Care must be taken when considering the GUP framework.
Indeed, in view of the commutation relation \eqref{waven} and the fact that \({k}\) and \({p}\) are no longer interchangeable, it becomes evident that
the role of spatial translations generator belongs to the wavenumber ${k}$ \cite{Bosso:2022rue}.
For instance, applying \({T}(x_0)\) to the state in Eq.~\eqref{Jack} in the case of a KMM deformation yields, in the momentum representation,
\begin{align}
\nonumber
  \hspace{-2.3mm}\phi_{\text{sat}}^{(x_0)}(p)
  &\hspace{-0.1mm}\equiv\hspace{-0.1mm} \langle p| T(x_0)|\phi_{\text{sat}}\rangle\\[2mm]
  &\hspace{-0.1mm}=\hspace{-0.1mm} c \left(1+\beta p^2\right)^{-\frac{3}{4} - \frac{1}{4\beta \Delta p^2_{\text{sat}}}} e^{-i x_0 \frac{\arctan\left(\sqrt{\beta}p\right)}{\hbar\sqrt{\beta}}},
  \label{phix0}
\end{align}
with $c$ given in Eq.~\eqref{norm}.

As a first check, one can verify that $\langle \phi_{\text{sat}}^{(x_{0})} |  x | \phi_{\text{sat}}^{(x_0)} \rangle = x_0$.
We notice that Eq.~\eqref{phix0} remains a solution of the differential equation~\eqref{GenDifEq}, with the first term replaced by $ x^2 \to (x - x_0)^2$, consistently with the prescription given in~\cite{Jackiw:1968zzb}.

\begin{figure}[t]
\centering\includegraphics[width=0.49\textwidth]{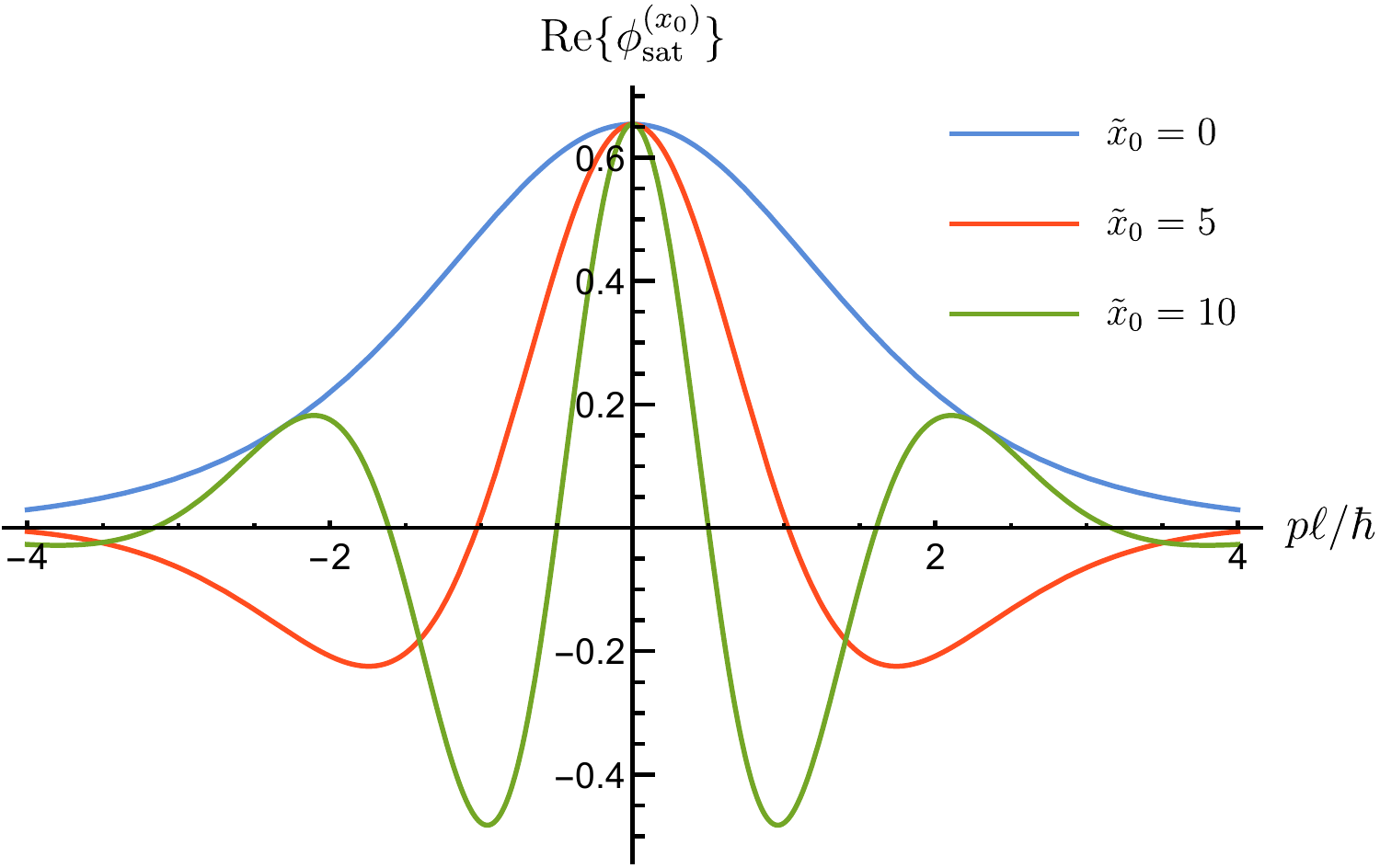}
\caption{Plot of $\Re\hspace{0.2mm}\{\phi_{\text{sat}}^{(x_0)}\}$ as a function of the dimensionless momentum $p\ell/\hbar$, for various values of $\tilde x_0=x_0/(\hbar\sqrt{\beta})$. We use the same unit convention as in Fig. \ref{Fig1} and set $\beta_0=0.1$.
}%
\label{x0}%
\end{figure}

The behavior of the real part of $\phi_{\text{sat}}^{(x_0)}(p)$ versus $p$ is shown in Fig.~\ref{x0} for various values of the dimensionless parameter $\tilde{x}_0 = x_0 / (\hbar \sqrt{\beta})$. The resulting functions remain peaked around $p = 0$, but due to the presence of an additional phase factor, they exhibit oscillations whose amplitude and frequency increase with increasing $\tilde{x}_0$.

\begin{figure}[t]
\centering\includegraphics[width=0.5\textwidth]{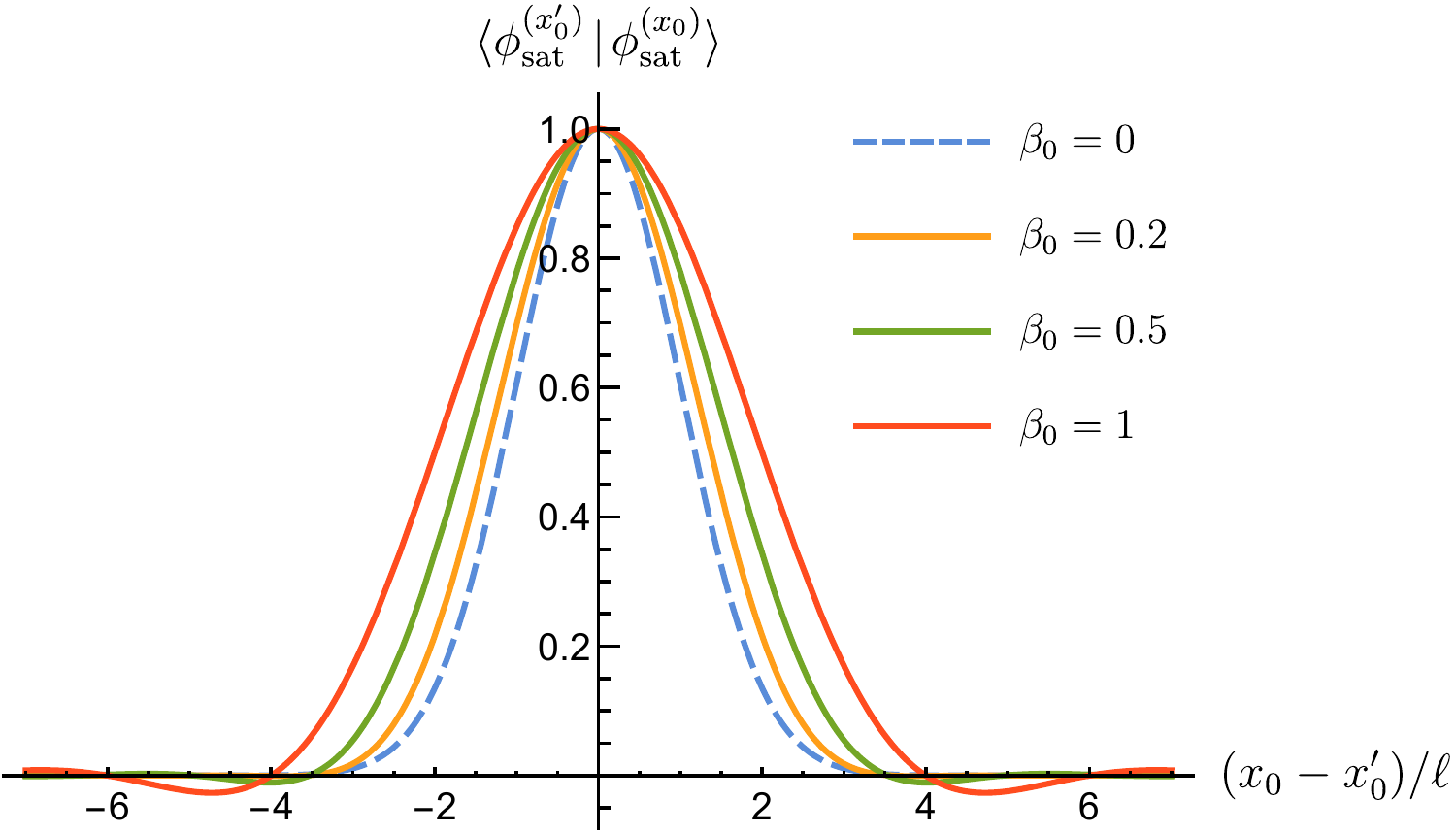}
\caption{Plot of the inner product \eqref{scalprod} versus the dimensionless scale $(x_0-x_0')/\ell$, for various values of the GUP parameter $\beta_0$. We use the same unit convention as in Fig. \ref{Fig1}.  
}%
\label{Scal}%
\end{figure}

We can also evaluate the scalar product between two states from the set in Eq.~\eqref{phix0} for different values of $x_0$, obtaining
\begin{equation}
    \label{scalprod}
\langle \phi_{\text{sat}}^{(x'_{0})} \hspace{0.2mm} | \hspace{0.2mm} \phi_{\text{sat}}^{(x_{0})}\rangle=\frac{\left\{\Gamma\left[\frac{1}{2}\left(3+\frac{1}{\beta\Delta p_{\text{sat}}^2}\right)\right]\right\}^2}{\Gamma_+ \Gamma_-}\,,
\end{equation}
where we have used the shorthand notation
\begin{equation}
\label{ort2}
    \Gamma_{\pm}\equiv\Gamma\left[\frac{1}{2}\left(3+\frac{1}{\beta\Delta p_{\text{sat}}^2}\pm\frac{x_0-x_0'}{\hbar\sqrt{\beta}}\right)\right].
\end{equation}
Therefore, the states in Eq.~\eqref{phix0} are generally non-orthogonal, similarly to their counterparts in the framework of QM. This can be clearly seen in Fig.~\ref{Scal}, which shows that, as the GUP parameter increases, the overlap between the states increases in a non-trivial way, as a consequence of the enhanced delocalization induced by the GUP.
The standard QM limit is consistently recovered for $\beta \to 0$.

Finally, following analogous reasoning and taking into account the commutation relation \eqref{canpq}, we observe that it is possible to boost the states~\eqref{GUPCSbetarid} in momentum space by applying the operator \({S}(p_0) = e^{i p_0 {q}/\hbar}\).

\section{Physical Interpretation and Dynamics}
\label{PIaD}

\subsection{Time Evolution and Classical Trajectories}

Having noticed that CSs, as defined in Section~\ref{MLQM}, are not minimal-uncertainty states, although most of the properties of CSs in ordinary QM are retained, it is now worth characterizing their physical properties. Specifically, given their definition, the standard features still hold when the auxiliary position $q$ and the momentum $p$ are considered. In particular, the time evolution of their expectation values is simply given by
\begin{align}
    \langle q \rangle(t)
    &= \sqrt{\frac{2 \hbar}{m \omega}} \Re \left[\alpha e^{- i \omega t}\right],\\[2mm]
    \langle p \rangle(t)
    &= \sqrt{2 \hbar m \omega} \, \Im \left[\alpha e^{- i \omega t}\right].
\end{align}
However, due to the non-trivial relation between $q$ and the physical position $x$, the latter exhibits new features.

\begin{figure}
    \centering
    \begin{subfigure}{0.5\textwidth}
        \includegraphics[width=\linewidth]{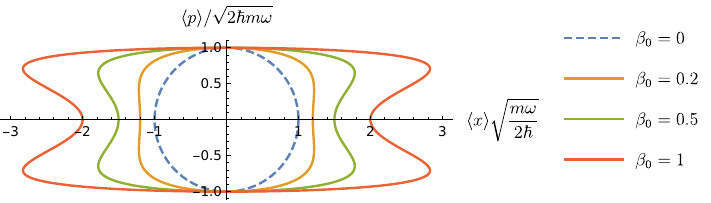}
        \caption{Quadrature-space trajectories of coherent states.}
        \label{subfig:CS_quadratures}
    \end{subfigure}
    \begin{subfigure}{0.48\textwidth}
        \includegraphics[width=\linewidth]{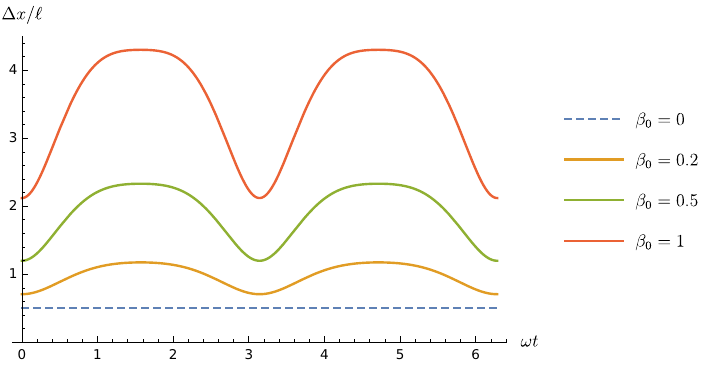}
        \caption{Evolution of the ratio $\Delta x(t)/\ell$ with time.}
        \label{subfig:CS_DeltaX}
    \end{subfigure}
    \caption{The plots above are for different values of the deformation parameter $\beta_0$, thus representing scenarios with different deformation strengths.
    For these plots, we used $\alpha = 1$ as well as $\Delta p = \sqrt{\frac{\hbar m \omega}{2}} = \hbar / \ell$.}
    \label{fig:CS}
\end{figure}

It is then convenient to move to the Heisenberg picture and write
\begin{align}
    q(t)
    &= \sqrt{\frac{\hbar}{2 m \omega}} \left[a e^{- i \omega t} + a^\dagger e^{i \omega t}\right],\\[2mm]
    p(t)
    &= - i \sqrt{\frac{\hbar m \omega}{2}} \left[a e^{- i \omega t} - a^\dagger e^{i \omega t}\right].
\end{align}
Then, using Eq.~\eqref{factord} with $f(p) = 1 + \beta p^2$, we obtain the results shown in Fig.~\ref{fig:CS}.
Specifically, we observe that the value of the position quadrature depends on the momentum quadrature, leading to a deformation of the phase-space trajectory whose magnitude is controlled by the strength of minimal-length effects. Furthermore, we find that the position uncertainty evolves periodically in time with period $\pi/\omega$, i.e. half the oscillation period. Both features clearly distinguish CSs in minimal-length models from their counterparts in ordinary QM.

We now compare these results with the case of a minimal-uncertainty state $|\phi_{\text{sat}}\rangle$. For convenience, we assume that $\langle x \rangle = 0$ and $\langle p \rangle = 0$.

\begin{figure}
    \centering
    \includegraphics[width=\linewidth]{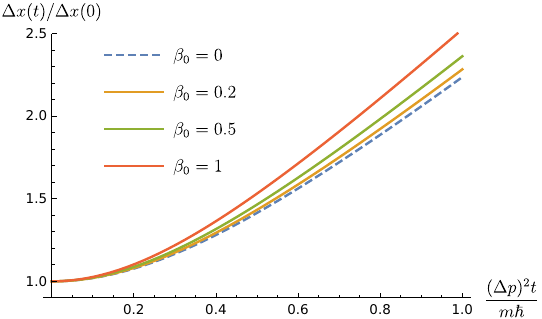}
    \caption{Dispersion of minimal uncertainty states.
    We observe that the larger the minimal length, that is the larger the parameter $\beta$, the faster the dispersion.
    For these plots, we have set $\Delta p = \frac{\hbar}{4 \ell}$, compatible with any values of $\beta$ considered in this plot, according to the observations at the end of Section \ref{Minimize}.}
    \label{fig:MinLenFree}
\end{figure}

Let us first consider the free evolution of such a state. Since the Hamiltonian depends only on the momentum operator, the Heisenberg evolution of the position operator is governed by
\begin{equation}
    x(t)
    = x + \frac{p t}{m} \left(1 + \beta p^2\right).
\end{equation}
In this case, we observe that the position uncertainty undergoes a dispersion that depends on the parameter $\beta$, or equivalently on the size of the minimal length. Specifically, the larger the minimal length, the slower the dispersion, as shown in Fig.~\ref{fig:MinLenFree}.

\begin{figure}
    \centering
    \begin{subfigure}{0.45\textwidth}
        \centering
        \includegraphics[width=\linewidth]{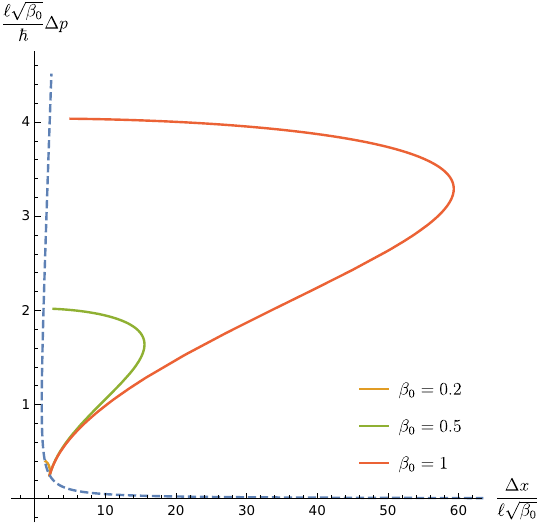}
        \caption{
        The dashed blue line represents the values of $\Delta x$ and $\Delta p$ saturating the generalized uncertainty principle.
        The solid lines correspond to the uncertainty profile of a maximally localized state as it evolves in a harmonic potential.}
        \label{subfig:MinLenHO_phase}
    \end{subfigure}
    \begin{subfigure}{0.48\textwidth}
        \centering
        \includegraphics[width=\linewidth]{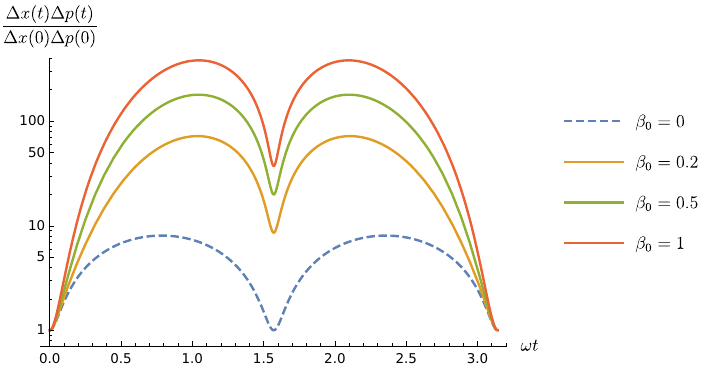}
        \caption{Evolution of the normalized uncertainty product for different values of the parameter $\beta_0$.}
        \label{subfig:MinLenHO_time}
    \end{subfigure}
    \caption{Uncertainty profile of a maximally localized state in a harmonic potential.
    Here, we considered states characterized by $\Delta p(0) = \frac{\hbar}{4\ell}$ and a harmonic potential such that $\sqrt{\frac{\hbar m \omega}{2}} = \hbar / \ell$.}
    \label{fig:MinLenHO}
\end{figure}

Finally, let us consider the case of a maximally localized state in the presence of a harmonic potential. This case is shown for different values of the parameter $\beta_0$ in Fig.~\ref{fig:MinLenHO}.
We observe that a state initially saturating the uncertainty relation gradually departs from the saturation condition and subsequently returns to it, with a period equal to half the harmonic oscillator period.
Furthermore, this ``breathing'' feature becomes more pronounced for larger values of the parameter $\beta_0$, and therefore for larger values of the minimal length.


Besides the time-dependent evolution of the uncertainty width, it is worth noticing that, unlike the ordinary case, a maximally localized state contains not only frequency components at the harmonic oscillator frequency $2\omega$, but also higher harmonics at frequencies $4\omega$ and $6\omega$. These additional modes originate from the non-linear term $\beta p^2$ entering the definition of the physical position operator. The resulting distortion of the uncertainty profile is therefore a genuinely quantum-gravitational effect induced by the presence of a minimal length.

From a phenomenological perspective, the appearance of higher harmonics could provide a signature of GUP-induced corrections. Indeed, 
the emergence of additional peaks in the Fourier spectrum of the uncertainty dynamics would signal departures from ordinary quantum theory. Such effects may become accessible in high-precision semiclassical systems, where tiny deviations from harmonic evolution can be amplified and spectrally resolved. In particular, optomechanical setups and analogue-gravity platforms have been proposed as promising scenarios to test Planck-scale modifications of the dynamics through precision frequency measurements and noise spectroscopy~\cite{Ali:2011fa,Bawaj:2014cda,Pikovski:2011zk}.

\subsection{The ``Squeezing'' Effect}

Here, it is worth commenting on the squeezing properties of the states $|\phi_{\text{sat}}\rangle$.
In the standard theory, squeezing acts as a dynamical resource: the uncertainties of the two conjugate variables $x$ and $p$ can be redistributed while keeping their product constant.
In the presence of a minimal length, however, CSs no longer saturate the uncertainty relation. Consequently, a state $|\phi_{\text{sat}}\rangle$ that does saturate the GUP necessarily appears squeezed relative to the standard vacuum state.

This squeezing effect induced by a minimal length becomes even more evident in the momentum-space representation of the saturating state $|\phi_{\text{sat}}\rangle$.
Specifically, as shown in Fig.~\ref{Fig1}, the larger the minimal length $\ell$, the narrower the momentum-space probability density.
As expected, in the position-space representation, this behavior is reversed: increasing the minimal length results in a broader position-space probability density (see Appendix \ref{AppB}).

We therefore observe a ``squeezing'' effect induced by the presence of a minimal length, in the sense that the momentum-space probability distribution becomes more sharply peaked than its counterpart in ordinary quantum theory. Conversely, the position-space distribution exhibits an ``anti-squeezing'' behavior, spreading over a wider spatial region as the minimal length increases.  As a result, the minimal length effectively acts as an intrinsic source of ``squeezing'', modifying the balance between position and momentum fluctuations relative to the standard QM scenario, even for states that saturate the GUP.

It is also worth emphasizing that, while in ordinary QM the position-space distribution can in principle be squeezed arbitrarily, this is no longer possible in the presence of a minimal length. Indeed, the latter introduces a fundamental lower bound on spatial localization, thereby imposing a natural cut-off on the amount of admissible position-space ``squeezing''. This observation is consistent with previous results \cite{Kempf:1994su,Bosso:2017ndq}, which showed that an infinite amount of energy is required to approach the absolute minimal position uncertainty allowed by the model.

\subsection{Energy Mean and Variance}

We now turn to the energetic characterization of the states in Eqs.~\eqref{AnnGUP} and \eqref{GUPCSbetarid}.
Concerning the former, it is easy to understand that, in terms of the harmonic-oscillator Hamiltonian \eqref{GUPHam}, the expectation value of the energy is given by Eq. \eqref{exvalHam}, which we rewrite here for convenience in the form 
\begin{equation}
\label{valHalpha}
\langle \mathcal{H}\rangle_{\alpha}
    = \hbar \omega \left( |\alpha|^2 + \frac{1}{2} \right). 
\end{equation} 
Assuming $\langle  q\rangle_{\alpha}=\langle  p\rangle_{\alpha}=0$, or equivalently $\Re\{\alpha\}=\Im\{\alpha\}=0$, this reduces to the standard vacuum energy
$\langle\mathcal{H}\rangle_{0}=\hbar\omega/2$.

Similarly, for the standard deviation one finds 
\begin{equation}
\label{deltaHalpha}
\Delta \mathcal{H}_\alpha=\hbar\omega|\alpha|\,,
\end{equation}
which trivially implies $\Delta \mathcal H_{0}=0$.
    
On the other hand, we recall that the states \eqref{GUPCSbetarid} are defined purely at the kinematical level through Eq.~\eqref{Jack}, without reference to any underlying Hamiltonian. However, in order to gain some insight into their energetic properties, we may first consider an effective free-particle description and evaluate the expectation value of the kinetic energy operator. In this case, we simply obtain
\begin{align}
 \langle \mathcal{H}^{(\mathrm{free})} \rangle_{\text {sat}}
&\equiv
\langle \phi_{\mathrm{sat}} \lvert \frac{p^{2}}{2m} \rvert \phi_{\mathrm{sat}} \rangle = \frac{\Delta p^2_{\text {sat}}}{2m}\,,\\[2mm]
\Delta \mathcal{H}^{\text{(free)}}_{\text{sat}}&=\frac{\Delta p^2_{\text {sat}}}{2m}\sqrt{\frac{2\left(1+\beta\Delta p^2_{\text {sat}}\right)}{1-2\beta\Delta p^2_{\text {sat}}}}\,,
\end{align}
where the second equality is valid provided that $\langle p^4\rangle$ is finite, which requires $2\beta \Delta p_{\text{sat}}^2 < 1$.
This is related to the issue discussed at the end of Sec.~\ref{Minimize}.
We note that the maximally localized states introduced in \cite{Kempf:1994su}, for which $\Delta p = 1/\sqrt{\beta}$, do not satisfy this condition. Accordingly, although they are proper physical states with finite mean energy $\langle \mathcal{H}^{(\mathrm{free})} \rangle_{\text{sat}} = 1/(2m\beta)$ \cite{Kempf:1994su}, they cannot be regarded as semiclassical energy states due to the divergence of the energy fluctuations. This behavior is not unexpected, since maximally localized states are constructed to optimize spatial localization rather than to achieve sharp energy localization. For comparison, recall that plane waves in momentum space (or Dirac $\delta$-functions in position space), which formally correspond to maximally localized states in ordinary quantum mechanics, have ill-defined expectation values for all powers of the momentum operator.

To allow for a more direct comparison with Eqs. \eqref{valHalpha} and \eqref{deltaHalpha}, we now regard $|\phi_{\text{sat}}\rangle$ as a trial state for the harmonic oscillator and characterize its properties with respect to the Hamiltonian \eqref{GUPHam}. We obtain
\begin{align}
   \hspace{-0.5mm} \langle \mathcal{H}^{\text{(ho)}} \rangle_{\text{sat}}
    &= \frac{\Delta p^2_{\text{sat}}}{2m}
    + \frac{\hbar^2 m \omega^2}{8 \Delta p^2_{\text{sat}}} \frac{\mathcal{F}_{1,2} \mathcal{F}_{1,3}}{\mathcal{F}_{1,5}},
    \label{Hsat}\\[2mm]
    \Delta \mathcal{H}^{\text{(ho)}}_{\text{sat}}
    &= \frac{1}{4} \left\{\frac{\hbar^4 m^2 \omega^4\,\mathcal F_{1,2}\mathcal F_{1,3}}{2\Delta p^4_{\text{sat}}\mathcal F_{1,5}^{\,2}\mathcal F_{1,9}}\right.\nonumber\\
    &\hspace{-8mm} \times \left[1+\beta\Delta p^2_{\text{sat}}\left(
    17+3\beta\Delta p^2_{\text{sat}}
    \mathcal F_{33,61}
    \right)
    \right]+
    \frac{
    8\Delta p^4_{\text{sat}}
    \mathcal F_{1,1}
    }{
    m^2\mathcal F_{1,-2}
    }
    \nonumber
    \\
    &\hspace{-8mm}\left.-
    \frac{
    4\hbar^2\omega^2
    \left(
    1+\beta\Delta p^2_{\text{sat}}
    \mathcal F_{8,3}
    \right)
    }{
    \mathcal F_{1,5}
    }
    \right\}^{1/2},
\end{align}
where the second equation is once again valid for \(2\beta \Delta p^2_{\text{sat}} < 1\), and we have defined for brevity
\begin{equation}
\mathcal{F}_{\gamma,\lambda}\!\left(\Delta p^2_{\text{sat}}\right)
    \equiv \gamma + \lambda\, \beta\, \Delta p^2_{\text{sat}} \, .
\end{equation}

In the limit $\beta\to0$, both the expectation value $\langle \mathcal{H}^{\text{(ho)}}\rangle_{\text{sat}}$ and the corresponding uncertainty $\Delta \mathcal{H}^{\text{(ho)}}_{\text{sat}}$ correctly reduce to the standard harmonic-oscillator results. In particular, upon imposing the ordinary QM relation $\Delta p_{\text{sat}}^2=\hbar m\omega/2$, one recovers the usual zero-point energy $\hbar\omega/2$, while the energy uncertainty vanishes. Indeed, in this regime, the momentum-space wave function \eqref{GUPCSbetarid} reduces to the familiar Gaussian ground-state profile.

For $\beta>0$, the situation changes significantly. The GUP-deformed saturating state \eqref{GUPCSbetarid} is no longer Gaussian and does not coincide with the harmonic-oscillator vacuum. As a consequence, both the mean energy and its fluctuation acquire nontrivial $\beta$-dependent corrections. In particular, the expectation value $\langle \mathcal{H}^{\text{(ho)}}\rangle_{\text{sat}}$ increases monotonically with $\beta$. This behavior originates from the GUP-induced deformation of the momentum-space profile of the state \eqref{GUPCSbetarid}, which enhances the expectation value of the canonical coordinate operator $q^2$ entering the harmonic potential. As a consequence, the potential-energy contribution increases, leading to a larger total mean energy of the oscillator.

At the same time, the energy uncertainty $\Delta \mathcal{H}^{\text{(ho)}}_{\text{sat}}$ becomes nonvanishing as soon as $\beta\neq0$. This reflects the fact that the GUP-saturating states are no longer stationary states of the harmonic oscillator Hamiltonian. Therefore, unlike the ordinary vacuum, the deformed states do not possess a sharply defined energy. The growth of $\Delta \mathcal{H}^{\text{(ho)}}_{\text{sat}}$ with increasing $\beta$ provides a quantitative measure of the departure from the standard harmonic-oscillator ground state and signals the emergence of intrinsically quantum-gravitational fluctuations induced by the minimal-length deformation.

More generally, the simultaneous increase of both $\langle \mathcal{H}^{\text{(ho)}}\rangle_{\text{sat}}$ and $\Delta \mathcal{H}^{\text{(ho)}}_{\text{sat}}$ highlights how the GUP deformation progressively drives the system away from the canonical coherent-vacuum configuration. In this sense, the parameter $\beta$ controls the strength of the deviation from ordinary QM: the larger $\beta$, the larger the deformation of the momentum distribution and the stronger the induced corrections to both the average energy and its quantum fluctuations.



\section{Conclusions and Outlook}
\label{Conc}

In this work, we have performed a systematic investigation of CSs for the one-dimensional harmonic oscillator within the framework of the quadratic GUP. Our analysis shows that the well-known equivalence among the standard characterizations of CSs in ordinary QM - namely, as eigenstates of the annihilation operator, displaced vacuum states and minimum-uncertainty wave packets - is generically lost once the Heisenberg algebra is deformed.

We remark that the origin of this breakdown can be traced back to the interplay between the auxiliary canonical coordinate $q$, satisfying $[q,p]=i\hbar$, and the physical position operator $x$, which obeys the deformed commutation relation \eqref{GUP}. Indeed, consistency with GUP-deformed Galilean invariance requires the harmonic-oscillator Hamiltonian to be formulated in terms of the canonical pair $(q,p)$, so that generalized ladder operators and algebraic CSs are naturally defined in the auxiliary canonical sector. 
On the other hand, the uncertainty relation \eqref{GUP} involves the pair $(x,p)$, since $x$ represents the physically relevant position operator. As a consequence, the states obtained as eigenstates of the generalized annihilation operator preserve the formal harmonic-oscillator structure in the $(q,p)$ sector, but do not, in general, saturate the physical GUP associated with $(x,p)$. Conversely, the states minimizing the generalized uncertainty relation exhibit intrinsically non-Gaussian momentum-space profiles and cannot be interpreted as ordinary displaced vacua. In this sense, the minimal-length deformation disentangles the notions of coherence and localization, which coincide exactly only in the undeformed limit $\beta\to0$.

The physical implications of the minimal-length framework turn out to be highly nontrivial. While the auxiliary canonical coordinate $q$ still follows the ordinary harmonic-oscillator evolution, the physical position operator $x$ acquires a nonlinear momentum dependence induced by the GUP. Consequently, the associated phase-space trajectories become distorted with respect to the canonical harmonic motion, with deviations whose magnitude is directly controlled by the deformation parameter $\beta$.

Our analysis also reveals the emergence of an intrinsic squeezing mechanism induced purely by the GUP. In ordinary QM, CSs preserve a constant uncertainty profile during time evolution. By contrast, within the GUP framework the physical uncertainty oscillates periodically, alternately approaching and departing from the minimum-uncertainty configuration. This phenomenon becomes increasingly pronounced as the deformation parameter $\beta$ grows, signaling the progressive departure from canonical semiclassical behavior.

A further important result concerns the energetic properties of the GUP-saturating states. While the ordinary harmonic-oscillator vacuum is recovered in the undeformed limit, for $\beta>0$ both the expectation value of the Hamiltonian and the corresponding energy uncertainty acquire nontrivial corrections. In particular, the mean energy increases with $\beta$, while the energy uncertainty becomes nonvanishing, showing that the GUP-saturating states are no longer exact stationary states of the harmonic oscillator. In this sense, the minimal-length deformation induces intrinsically quantum-gravitational fluctuations even in states that reduce to the ordinary vacuum configuration as $\beta\to0$.

Possible future developments naturally emerge from the present analysis. A first direction concerns the role of generalized CSs in the quantum-to-classical transition and in gravity-induced decoherence mechanisms. Since CSs represent the most robust semiclassical configurations in ordinary QM, it would be particularly interesting to investigate how minimal-length deformations modify the emergence of classicality in semiclassical gravity frameworks. In particular, the intrinsically nonvanishing energy fluctuations and the squeezing induced by the GUP suggest that the corresponding states may undergo decoherence differently from ordinary Gaussian CSs, potentially affecting the stability of semiclassical spacetime backgrounds, cosmological perturbations and inflationary quantum fluctuations. Such aspects may provide new insight into the longstanding problem of how classical spacetime emerges from an underlying quantum-gravitational regime \cite{Kiefer:1998qe}.

A second promising direction concerns the interplay between generalized CSs and emergent gravitational phenomena in condensed-matter analog systems. In particular, graphene and related Dirac materials provide especially intriguing platforms, since effective GUP-like structures can naturally arise from the underlying lattice geometry and quasiparticle dynamics~\cite{Iorio:2022ave}. In this framework, departures from the standard Dirac regime induce modified commutation structures and nontrivial semiclassical dynamics. Consequently, the non-Gaussian CSs and dynamical squeezing effects analyzed in the present work may influence wave-packet propagation and coherence properties of low-energy excitations, offering observable signatures of minimal-length-inspired physics in analog-gravity and emergent-spacetime scenarios.

Work along these directions is currently underway and will be presented elsewhere in future investigations.

\appendix

\section{Momentum-space representation}
\label{Momrepres}

In this appendix, we aim to solve Eq. \eqref{Jack} in momentum space.
We will first solve it in $k$-space and then transform it in $p$-space.
To do so, we need the $k$-space representation of the position operator $x$.
Thus, starting from Eq. \eqref{sympre}, we find
\begin{equation}
    x
    = \frac{i}{2} \left\{g(k), \frac{1}{g(k)} \frac{d}{dk}\right\},
\end{equation}
with $g(k) = f(p(k))$.
Thus, Eq. \eqref{Jack} can be in the equivalent form
\begin{equation}
    \left[\frac{i}{\Delta x_{\text{sat}}} \left(\frac{d}{dk} + \frac{1}{2 g(k)} \frac{d g(k)}{dk}\right) - \frac{i}{\Delta p_{\text{sat}}} p(k) \right] \tilde{\phi}_k(k) = 0\,,
\end{equation}
where the auxiliary function $\tilde{\phi}_k$ is defined as the result of the action of 
the second (right) operator in Eq.~\eqref{Jack} on $|\phi_{\text{sat}}\rangle$, 
while the operator acting on $\tilde{\phi}_k$ corresponds to the first (left) 
operator in Eq.~\eqref{Jack}, both of which, for convenience, are expressed 
using the wavenumber $k$ defined in Eq.~\eqref{waven}.
Solving this equation yields
\begin{equation}
    \tilde{\phi}_k(k)
    = \frac{c_1}{\sqrt{g(k')}} \exp \left[\frac{\Delta x_{\text{sat}}}{\Delta p_{\text{sat}}} \int_0^k p(k') \, d k'\right],
\end{equation}
where $c_1$ is an integration constant.

Recalling the definition of $\tilde\phi_k$, we can now write
\begin{multline}
     \phi_{\text{sat},k}(k)
    = \frac{\exp \left[- \frac{\Delta x_{\text{sat}}}{\Delta p_{\text{sat}}} \int_0^k p(k') \, dk'\right]}{\sqrt{g(k)}} \\[2mm]
    \times \left\{c_2
    - i c_1 \Delta x_{\text{sat}}
    \int_0^k \exp \left[\frac{2 \Delta x_{\text{sat}}}{\Delta p_{\text{sat}}} \int_0^{k'} p(k'') \, dk''\right] \, dk'\right\},
\end{multline}
with $c_2$ a second integration constant.
We now observe that the term multiplying $c_1$ is ill-behaved in the limit to ordinary QM.
Specifically, since in such a limit $p(k) = k$, the innermost integral diverges for large values of $k$.
Therefore, in order to obtain the correct asymptotic behavior, we will set $c_1=0$.
Consequently, we can write
\begin{equation}
    \phi_{\text{sat},k}(k)
    = \frac{c_2}{\sqrt{g(k)}}\exp \left[- \frac{\Delta x_{\text{sat}}}{\Delta p_{\text{sat}}} \int_0^k p(k') \, dk'\right].
    \label{eqn:wf_k}
\end{equation}
In this sense, $c_2$ acquires the meaning of normalization constant.
As for the corresponding wavefunction in $p$-space, we straightforwardly find
\begin{equation}
    \phi_{\text{sat}}(p)
    = \frac{c_2}{\sqrt{f(p)}} \exp \left[- \frac{\Delta x_{\text{sat}}}{\Delta p_{\text{sat}}} \int_0^p \frac{p'}{\hbar f(p')} \, dp'\right].
\end{equation}

\section{Basis Change via Fourier Transform}
\label{AppA}

In this appendix, we derive the Fourier transform connecting position and wavenumber eigenstates, highlighting the relation between the \( x \)-space and \( k \)-space representations.
In order to determine the appropriate transformation, we must first identify the eigenfunctions of the position operator $x$ expressed in $k$-space, namely \( \langle {k} | {x} \rangle \). Following~\cite{Bosso:2024nmn}, it can be shown that these functions satisfy the following equation:
\begin{equation}
\label{Geq}
    x\langle k | x \rangle=i\frac{\partial}{\partial k}\langle k|x\rangle + \frac{i}{2}\frac{\partial \log g(k)}{\partial k}\langle k|x\rangle\,.
\end{equation}
For the KMM model, the factor $g(k)$ takes the form~\cite{Bosso:2024nmn}
\begin{equation}
    g(k) = \left(1+\beta p^2(k)\right)= \sec^2\left(\hbar\sqrt{\beta}k\right).
    \label{G}
\end{equation}
We then obtain 
\begin{equation}
\label{Fourkx}
\langle k|x\rangle=\frac{\mathcal{N}_k}{\sqrt{g(k)}} e^{-i x k}\,,
\end{equation}
where the normalization is given by $\mathcal{N}_k=1/\sqrt{\hbar\hspace{0.2mm} V_\kappa}$, with $V_\kappa \equiv \int_\kappa dk = \frac{\pi}{\hbar \sqrt{\beta}}$ denoting the total volume of $k$-space

The above relation can be employed to evaluate the inner product between position eigenstates. Specifically, using the resolution of the identity in terms of $k$-eigenstates, namely $\mathbb{1}=\hbar \int_\kappa g(k) |k\rangle\langle k| dk$, one obtains
\begin{equation}
    \langle x| x'\rangle=\frac{1}{V_\kappa}\int_\kappa e^{i\left(x-x'\right)k}dk\,.
\end{equation}
Additionally, we can consider the relation
\begin{equation}
    \langle k | \left[\int_{-\infty}^\infty |x\rangle\langle x| \hspace{0.2mm} dx\right]|k'\rangle=\frac{2\pi}{V_\kappa}\langle k|k'\rangle\,,
\end{equation}
where we have used Eq.~\eqref{Fourkx} along with the orthonormality condition $\langle k|k'\rangle=\delta(k-k')/(\hbar g(k))$. Thus, we can read
\begin{equation}
    \int_{-\infty}^\infty |x\rangle\langle x| \hspace{0.2mm} dx=\frac{2\pi}{V_\kappa}\,.
\end{equation}
This relation allows us to determine the transformation from position to wavenumber space. In particular, for a given state $|\psi\rangle$, we have
\begin{equation}
\label{kpsTr}
    \phi(k)\equiv\langle k|\psi\rangle=\frac{1}{2\pi}\sqrt{\frac{V_\kappa}{\hbar g(k)}}\int_{-\infty}^\infty e^{-ixk}\langle x|\psi\rangle\, dx\,.
\end{equation}
In turn, the inverse transform is given by
\begin{equation}
\label{FTr}
    \psi(x)\equiv\langle x|\psi\rangle=\sqrt{\frac{\hbar}{V_\kappa}}\int_\kappa \sqrt{g(k)}\hspace{0.2mm} e^{ixk} \langle k|\psi\rangle\hspace{0.3mm} dk\,.
\end{equation}

Finally, we observe that, following similar considerations, it is likewise possible to explicitly compute the Fourier transform between momentum $p$ and position $x$ (see~\cite{Bosso:2024nmn} for further details).

\section{$x$-representation of GUP-saturating states and coherent states}
\label{AppB}

In Sec.~\ref{MLQM} we determined the GUP-saturating states and the CSs in the momentum representation. In order to gain a clearer physical insight into minimal-length effects, it is instructive to investigate how these states appear in the representation of the physical position $x$. Concerning the state $|\phi_{\text{sat}}\rangle$ in Eq.~\eqref{GUPCSbetarid}, we observe that, due to the modified commutator~\eqref{GUP}, the Fourier transform must accordingly be modified if one wishes to directly connect the $x$- and $p$-representations~\cite{Bosso:2021koi,Bosso:2024nmn}.
Alternatively, we can take advantage of the canonical commutation relation \eqref{waven} and transform the wave function in Eq. \eqref{eqn:wf_k} instead.
Thus, using Eq. \eqref{FTr} in Appendix \ref{AppA}, we can write \cite{Bosso:2024nmn}
\begin{equation}
\label{xpsisat}
    \psi_{\text{sat}}(x)
    = \sqrt{\frac{\hbar}{V_\kappa}}\int_{-\kappa}^\kappa \sqrt{g(k)} e^{i x k} \phi_{\text{sat},k}(k) ~ d k\,,
\end{equation}
where $g(k)=f(p(k))$ corresponds to the Jacobian of the transformation between the variables \( p \) and \( k \). 

Considering the case of the KMM model, with $\kappa = \frac{\pi}{2\hbar\sqrt{\beta}}$ and $g(k)$ given in Eq. \eqref{G}, 
we get
\begin{multline}
    \hspace{-4mm}\psi_{\text{sat}}(x)
    = \frac{2^{-\frac{1}{2} \left(1 + \frac{1}{\beta \Delta p_{\text{sat}}^2}\right)}}{{\pi^{3/4}}}\,\cos\left[\frac{\pi}{4}\left(3-\frac{2x}{\hbar\sqrt{\beta}}+\frac{1}{\beta\Delta p_{\text{sat}}^2}\right)\right]\\[2mm]
    \times\,\frac{\Gamma\left[\frac{1}{4}\left(-1+\frac{2x}{\hbar\sqrt{\beta}}-\frac{1}{\beta\Delta p_{\text{sat}}^2}\right)\right] \left\{{\Gamma\left[\frac{1}{2}\left(3+\frac{1}{\beta\Delta p_{\text{sat}}^2}\right)\right]}\right\}^{\frac{3}{2}}}
    {\Gamma\left[\frac{1}{4}\left(5+\frac{2x}{\hbar\sqrt{\beta}}+\frac{1}{\beta\Delta p_{\text{sat}}^2}\right)
    \right]\sqrt{\Gamma\left(1+\frac{1}{2\beta\Delta p_{\text{sat}}^2}\right)}}\,,
    \label{psisatx}
\end{multline}
where $\Delta p_{\text{sat}}$ can be rewritten as a function of $\Delta x_{\text{sat}}$ via Eq.~\eqref{dxsat}.

\begin{figure}[t]
\centering\includegraphics[width=0.45\textwidth]{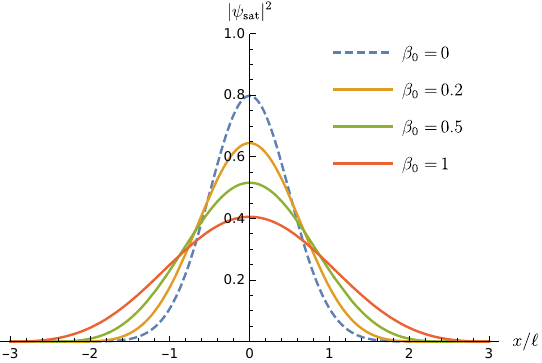}
\caption{Plot of $|\psi_{\text {sat}}^2|$ as a function of the dimensionless scale $x/\ell$, for various values of the GUP parameter $\beta_0$. We set $\Delta p = \hbar / \ell$.}
\label{Fig2}%
\end{figure}

It should be noted that, due to the specific definitions of the Fourier transform and its inverse in Eqs.~\eqref{kpsTr} and~\eqref{FTr}, one must multiply Eq.~\eqref{psisatx} by $\sqrt{V_\kappa / 2\pi}$ in order to obtain a properly normalized wavefunction in $x$-space, i.e., 
\begin{equation}
\label{B1}
    \psi^{\text{norm}}_{\text{sat}}(x)=\sqrt{\frac{V_\kappa}{2\pi}}\psi_{\text {sat}}(x)\,.
\end{equation}
The behavior of $|\psi^{\text{norm}}_{\text{sat}}(x)|^2$ as a function of $x$ is shown in Fig.~\ref{Fig2} for various values of the GUP parameter $\beta$.  
Consistently with the trend displayed in Fig.~\ref{Fig1}, whereby the momentum-space probability density 
$|\phi_{\mathrm{sat}}(p)|^2$ becomes increasingly peaked around the origin as the GUP parameter $\beta$ increases, 
the corresponding probability density in the physical position representation exhibits an increasing spatial 
broadening around $x=0$. On the other hand, in the limit $\beta \to 0$, the standard Gaussian behavior of ordinary QM (black dashed line), given by Eq.~\eqref{Gausx}, is correctly recovered. 

In order to further demonstrate analytically the consistency of this limit, we focus on the case in which the expectation values of position and momentum vanish. Similar considerations can be straightforwardly extended to the more general situation in which these quantities are non-vanishing.
To this end, we note that the argument of the \(x\)-independent Gamma functions in Eq.~\eqref{psisatx} is always positive and diverges in the limit \(\beta \rightarrow 0\). On the other hand, the argument of the \(x\)-dependent Gamma functions can be either largely negative or positive, depending on the value of \(x\). Therefore, we must consider both asymptotic expansions of the Gamma function:
\begin{align}
\label{exp1}
    \Gamma(y)&\overset{y \to +\infty}{\sim} \sqrt{\frac{2\pi}{y}}\left(\frac{y}{e}\right)^y\,,\\[2mm]
    \Gamma(y)&\overset{y \to -\infty}{\sim} -\dfrac{\pi}{\sin\left(\pi |y|\right)}\dfrac{1}{\sqrt{2\pi |y|}}\left(\dfrac{e}{|y|}\right)^{|y|}\,,
    \label{exp2}
\end{align}
with the latter following from Euler's reflection formula $\Gamma(y)=\dfrac{\pi}{\sin\left(\pi y\right)\Gamma(1-y)}$, valid for large negative values of \(y\).

For computational convenience, and considering that \(\psi_{\text{sat}}(x)\) is significantly non-zero only in a narrow region around \(x = 0\) (see Fig.~\ref{Fig2}), we restrict our analysis to values of \(x\) that are sufficiently small in absolute value, such that the argument of the Gamma function in the numerator of Eq.~\eqref{psisatx} is negative, while that in the denominator is positive (formally similar considerations can be applied if this condition is not fulfilled). Under these assumptions, by substituting the expansions~\eqref{exp1}-\eqref{exp2} into Eq.~\eqref{B1} and performing some algebraic manipulations, we obtain
\begin{widetext}
\begin{align}
\nonumber
     \psi^{\text{norm}}_{\text{sat}}(x) &\overset{\beta \to 0}{\sim} \hbar^{1+\frac{1}{2\beta\Delta p^2_{\text{sat}}}}\left(\frac{2\Delta p^2_{\text{sat}}}{\pi e}\right)^{\frac{1}{4}}\left(1+2\beta\Delta p^2_{\text{sat}}\right)^{-\frac{1}{4}\big(1+\frac{1}{\beta\Delta p^2_{\text{sat}}}\big)}
     \left(1+3\beta\Delta p^2_{\text{sat}}\right)^{\frac{3}{4}\big(2+\frac{1}{\beta\Delta p^2_{\text{sat}}}\big)}\\[2mm]
     &\hspace{-6mm}\times\left[\hbar\left(1+\beta\Delta p^2_{\text{sat}}\right)-2\sqrt{\beta}\Delta p^2_{\text{sat}}x\right]^{-\frac{1}{4}\big(3-\frac{2x}{\hbar\sqrt{\beta}}+\frac{1}{\beta\Delta p^2_{\text{sat}}}\big)}
     \left[\hbar\left(1+5\beta\Delta p^2_{\text{sat}}\right)+2\sqrt{\beta}\Delta p^2_{\text{sat}}x\right]^{-\frac{1}{4}\big(3+\frac{2x}{\hbar\sqrt{\beta}}+\frac{1}{\beta\Delta p^2_{\text{sat}}}\big)}.
\end{align}
\end{widetext}
In the limit as $\beta\rightarrow0$, we finally acquire
\begin{equation}
    \lim_{\beta\rightarrow0}\psi_{\text{sat}}^{\text{norm}}(x)=\left(\frac{2\Delta p^2_{\text{sat}}}{\pi\hbar^2}\right)^{\frac{1}{4}}\,e^{-\big(\frac{x\hspace{0.2mm}\Delta p_{\text{sat}}}{\hbar}\big)^2}\,,
\end{equation}
which coincides with Eq.~\eqref{psialphafinal}, using the definition~\eqref{Dxalpha} of $\Delta p_{\text{sat}}$ and assuming that the expectation values of \( {p} \) and \( {x} \) are zero.

\subsection{Coherent states}

For completeness, we also exhibit the $x$-representation of the CS \eqref{infsupGUP}. For the KMM model,  using the relation~\eqref{pkrel} together with the Fourier transform~\eqref{FTr}, 
and applying the same strategy as for Eq. \eqref{xpsisat},  we obtain the normalized wavefunction
\begin{equation}
    \psi_{\alpha}(x)= C\, 
G^{2,1}_{1,2} \Biggl( a \;\Bigg|\; 
\begin{matrix} 1 - \frac{b\,x}{2} \\0, \frac{1}{2} \; \end{matrix}
 \Biggr)\,,
\end{equation}
where $G^{2,1}_{1,2}$ is the Meijer G-function \cite{abramowitz1972handbook} and we have defined 
$C\equiv\dfrac{1}{\left(4\beta^2\pi^3\hbar^3m\omega\right)^{\frac{1}{4}}}$, $a\equiv \dfrac{1}{4\beta \Delta p_{\text{sat}}^2}$, $b=\dfrac{1}{\hbar\sqrt{\beta}}$.
For consistency with Eq. \eqref{psisatx}, we have here set the expectation values of position and momentum to zero. Considering the asymptotic behavior of the Meijer G-function, it is possible to verify that the result in Eq. \eqref{psialphafinal} is correctly recovered in the limit $\beta \rightarrow 0$.


\section*{Acknowledgments}
The research of GGL is supported by the postdoctoral fellowship program of the 
University of Lleida.
PB and GGL gratefully acknowledge the contribution of the LISA Cosmology Working Group (CosWG), as well as support from the COST Action CA23130, \textit{Bridging 
high and low energies in search of quantum gravity (BridgeQG)}.
Furthermore, GGL acknowledges the support from the COST Actions 
CA21136 - \textit{Addressing observational tensions in cosmology with 
systematics and fundamental physics (CosmoVerse)} and CA21106 -  
\textit{COSMIC WISPers in the Dark Universe: Theory, astrophysics and 
experiments (CosmicWISPers)}.

\bibliographystyle{apsrev4-1}
\bibliography{references}

\end{document}